\documentclass[aps, prb, reprint, superscriptaddress]{revtex4-1}
\pdfoutput=1

\usepackage{amsmath}
\usepackage{graphicx}
\usepackage{xcolor}

\newcommand{\alaloxal}{Al--AlO$_\mathrm{x}$--Al}
\newcommand{\alox}{AlO$_\mathrm{x}$}
\newcommand{\alalox}{Al--AlO$_\mathrm{x}$}
\newcommand{\gcm}{g~cm$^{-3}$}
\newcommand{\corundum}{Al$_2$O$_3$}

\begin{document}

\title{Simulating the fabrication of aluminium oxide tunnel junctions}
\date{\today}

\author{M. J. Cyster}
\affiliation{Chemical and Quantum Physics, School of Science, RMIT University, Melbourne, Australia}

\author{J. S. Smith}
\affiliation{Chemical and Quantum Physics, School of Science, RMIT University, Melbourne, Australia}

\author{N. Vogt}
\affiliation{Chemical and Quantum Physics, School of Science, RMIT University, Melbourne, Australia}
\affiliation{HQS Quantum Simulations GmbH, 76187 Karlsruhe, Germany}

\author{G.~Opletal}
\affiliation{Data61, CSIRO, Door 34 Goods Shed, Village St, Docklands VIC 3008 Australia}

\author{S. P. Russo}
\affiliation{Chemical and Quantum Physics, School of Science, RMIT University, Melbourne, Australia}

\author{J. H. Cole}
\email[Correspondence: ]{jared.cole@rmit.edu.au}
\affiliation{Chemical and Quantum Physics, School of Science, RMIT University, Melbourne, Australia}

\begin{abstract}
    Aluminium oxide (\alox) tunnel junctions are important components in a range of nanoelectric devices including superconducting qubits where they can be used as Josephson junctions.
    While many improvements in the reproducibility and reliability of qubits have been made possible through new circuit designs, there are still knowledge gaps in the relevant materials science.
    A better understanding of how fabrication conditions affect the density, uniformity, and elemental composition of the oxide barrier may lead to the development of lower noise and more reliable nanoelectronics and quantum computers.
    In this paper we use molecular dynamics to develop models of \alaloxal\ junctions by iteratively growing the structures with sequential calculations.
    With this approach we can see how the surface oxide grows and changes during the oxidation simulation.
    Dynamic processes such as the evolution of a charge gradient across the oxide, the formation of holes in the oxide layer, and changes between amorphous and semi-crystalline phases are observed.
    Our results are widely in agreement with previous work including reported oxide densities, self-limiting of the oxidation, and increased crystallinity as the simulation temperature is raised.
    The encapsulation of the oxide with metal evaporation is also studied atom by atom.
    Low density regions at the metal--oxide interfaces are a common feature in the final junction structures which persists for different oxidation parameters, empirical potentials, and crystal orientations of the aluminium substrate.
\end{abstract}

\maketitle

\section{Introduction}

Superconducting quantum computers often use aluminium oxide tunnel junctions as Josephson junctions to introduce the required nonlinearity.\cite{Makhlin2001,Wendin2007,Clarke2008,Martinis2009a,Wu2013,Wang2015,Muller2017,Nersisyan2019}
The tunnel barrier in such junctions is formed by a thin dielectric film of amorphous aluminium oxide (\alox) which separates two metallic contacts.
As interest has expanded in superconducting quantum computing architectures so too has the the importance of clarifying the materials science which governs the formation and stability of thin \alox\ films.
Understanding the microscopic details of the oxide layer is a present focus for identifying and mitigating noise sources in a range of superconducting electronic devices.\cite{Muller2017}

High quality trilayer \alaloxal\ tunnel junctions are most commonly produced using the double-angle evaporation process pioneered by \citeauthor{Dolan1977}.\cite{Dolan1977}
The aluminium layers are deposited through a lithographic mask at different angles to a substrate with an intervening low pressure oxidation which forms the oxide barrier.\cite{Roddatis2011}
Other fabrication methods which modify or even remove the standard Dolan bridge structure also employ a low pressure oxidation step.\cite{Satoh2015,Lecocq2011,Zhang2017}

\begin{figure*}
    \includegraphics[trim=0 18mm 0 0, clip, width=7 in]{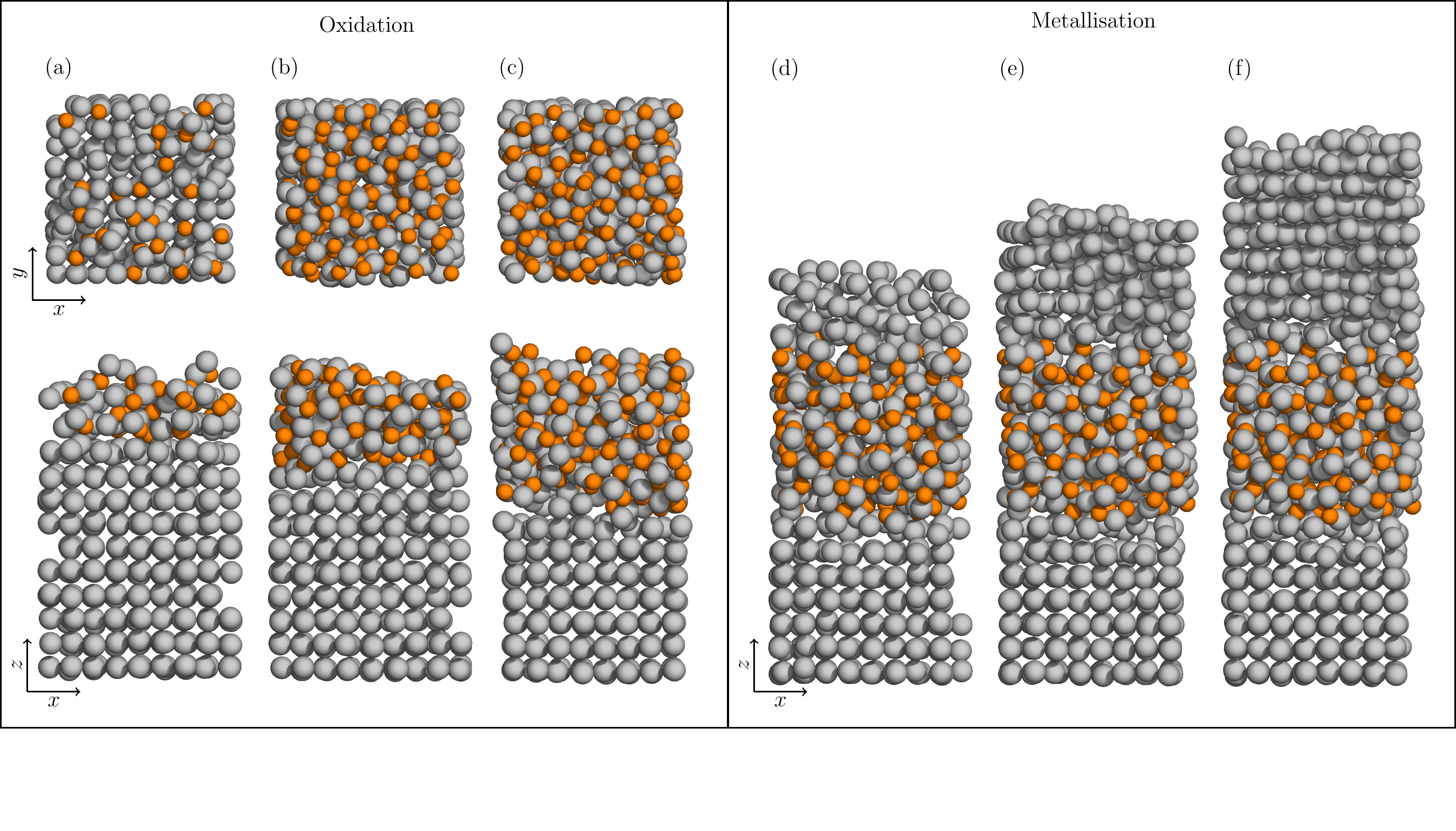}
    \caption
    {
        An amorphous oxide is formed on an Al(100) surface as oxygen atoms are iteratively added to the simulation cell.
        Oxygen and aluminium atoms are shown as orange and gray spheres respectively.
        (a)--(c) Views along and perpendicular to the $z$-axis after 30, 90, and 150 oxygen atoms have been introduced.
        (d)--(f) The growth of the second aluminium contact during the metallisation simulation.
        This calculation was performed with the S-M potential.
    }
    \label{fig:oxidation}
\end{figure*}

In this study we use molecular dynamics (MD) to explicitly model the low pressure oxidation and aluminium evaporation processes.
The structure of the oxide and junction emerge as oxygen and aluminium atoms are consecutively added to the surface.
In Fig.~\ref{fig:oxidation}(a)--(c) three stages of oxide growth are shown for a typical simulation.
The aluminium surface is partly then completely covered with oxygen as more atoms are deposited.
After the oxide layer is formed aluminium is added (metallisation) to completes the tri-layer junction structure [Fig.~\ref{fig:oxidation}(d)--(f)].

In Sec.~\ref{sec:oxidation} we outline the methodology we have used to simulate the oxidation process.
Structural properties such as density and stoichiometry are studied over the course of the simulated oxidation.
We also study how the charges on the atoms change as the dielectric barrier layer is formed and investigate the effect of temperature on the structure of the oxide.
Our results are then discussed in comparison to computational and experimental results from the literature.
In Sec.~\ref{sec:junction} we consider aluminium deposition onto a formed oxide layer of similar thickness to experimental reports\cite{Zeng2015a} (1.4--1.6~nm) to simulate the growth of \alaloxal\ junctions.
We examine how the structure of these junctions changes as they grow and make comparisons between the two empirical potentials we have used in this work.

\section{Oxidation of aluminium}
\label{sec:oxidation}

\subsection{Methodology}

One standard approach to studying oxidation computationally is to first create a region of aluminium surrounded by vacuum space before filling the vacuum with a nominal density of oxygen atoms or molecules.
The system is then allowed to evolve until a stable oxide layer forms on the aluminium surface.
In such studies the oxygen gas density frequently corresponds to an unrealistically high pressure ($\sim$10--500~atm) in order to accelerate the dynamics and reduce the required computational resources.\cite{Campbell2005,Zhou2005,Hasnaoui2006,Zhou2007,Hong2015} 
By comparison experimental junction fabrication is normally performed under high or ultra-high vacuum and partial oxygen pressures can vary over many orders of magnitude from $10^{-12}$ to $10^{-2}$~atm.\cite{Tan2005,Holmqvist2008,Cai2011,Wang2015}
Another way to develop models of \alaloxal\ junctions is to use a simulated annealing approach. 
In this case a crystalline \corundum\ structure is simulated at a temperature above the melting point to generate disorder before being cooled to lock the atoms into a particular configuration.\cite{Gutierrez2000,Colleoni2015,DuBois2016}

As an alternative to artificially raising the gas pressure or creating disorder with annealing we simulate the oxidation process directly by iteratively adding atoms to the surface.
We use MD to model the approach and bonding of individual oxygen atoms to the bare aluminium surface.
As we model individual atoms approaching the surface, we need not simulate the relatively long periods when no atoms are interacting with the surface. 
This results in a considerable reduction in the computational cost and corresponds to the high vacuum limit. 
A derivation is provided in Appendix~\ref{appendix:gas_surface} which supports this approximation.

While experimentally molecular oxygen (O$_2$),\cite{Su2017,Wu2013} ozone (O$_3$),\cite{Park2002} and both charged (O$^{+}$)\cite{Suzuki1993,Roos2001} and neutral atomic oxygen (O)\cite{Schuhl1990,Kleiman2003} can be used, it is known that O$_2$ has a large dissociation energy which the empirical potentials have not been designed to describe.\cite{Brune1992}
\citeauthor{Campbell2005} simulated the oxidation of nanoclusters and found similar behaviour for both neutral atomic oxygen and O$_2$ molecules excepting a change in the temperature at the surface.\cite{Campbell2005}
We therefore deposit individual neutral oxygen atoms in the present work for simplicity (except where otherwise stated).

Most of the calculations in this work were performed with the General Lattice Utility Program (GULP).\cite{Gale2003}
This program includes an empirical potential developed by \citeauthor{Streitz1994} (S-M) which describes the interactions between aluminium and oxygen atoms.\cite{Streitz1994}
As the aluminium oxide is an ionic material, charge transfer between atoms is an important component of this potential.
We solve the equations of motion and the distribution of charge every 1~fs.
The system is simulated as an NVT ensemble held at the chosen temperature by using the Nos\`e--Hoover thermostat.\cite{Nose1984,Hoover1985}
The coupling between the system and the heat bath is an important parameter of the thermostat. 
A detailed description of how we choose this value for various simulations is provided in Appendix~\ref{appendix:thermostat}.

For comparison we have also performed MD with LAMMPS.\cite{Plimpton1995,LAMMPS}
The parameters of these simulations (temperature, timestep, duration, etc.)~are identical to those we use in GULP.
Interactions between atoms -- including the charge equilibration processes -- are described by the ReaxFF force field\cite{vanDuin2001,Aktulga2012} using parameters for aluminium and oxygen published by \citeauthor{Hong2015}.\cite{Hong2015}

Aluminium substrates are prepared by creating supercells with the experimentally reported lattice constant of 4.041386~\AA.\cite{Cooper1962}
An optimisation of the geometry is performed in the MD software (either GULP or LAMMPS) where the atomic positions and the dimensions of the supercell are allowed to change to find the lowest energy configuration.
Vacuum is then added to increase the $z$ dimension of the supercell to 20~nm before a second optimisation which allows for aluminium layers to expand at the metal--vacuum interfaces.
This is the direction in which the oxide will grow as oxygen atoms are deposited.
The lattice constant of substrates optimised with S-M and ReaxFF potentials differ very slightly however both are within $\pm$0.2\% of the experimental value.

There are two steps in our methodology for each atom added to the surface, each involving a different MD simulation.
First the atom is positioned 2.4~nm from the existing surface with a randomised position in $x$ and $y$.
The initial velocity is obtained from the Maxwell--Boltzmann distribution and constrained to be directed towards the surface.
The system is then allowed to evolve for 15~ps.
If the atom has been reflected from the surface or is not bonded for any other reason the iteration is discarded and a new atom added. 
If the atom is bonded then a relaxation calculation is performed where the system equilibrates for 2~ps.
This technique serves to separate the individual atomic depositions so that they can be considered to be independent events.
In order to simulate the addition of enough atoms to form the surface oxide and the second electrode -- approximately 300 atoms for a substrate with a side length of $x = y =16$~\AA\ -- a total simulation time of 5~ns or more is required.
While this forms a substantial computational challenge, it is many orders of magnitude shorter than the minutes of oxidation in experiments.

\subsection{Results}

From the positions of the atoms we calculate the material density as a function of $z$ (the direction of the oxide growth).
Density is reported in units of $\rho$ where $\rho = 1$ corresponds to the density of crystalline \corundum\ (3.97~\gcm).\cite{CRC}
This is achieved by taking a Gaussian window with a full-width half-maximum (FWHM) of 2.4~\AA\ ($\sigma \simeq 1$~\AA) and moving it along $z$ in increments of 0.05~\AA.
The FWHM of the Gaussian is taken from the position in the radial distribution function $g(r)$ between the first and second peaks.
This same distance is used to determine coordination numbers and reflects the average distance between nearest neighbour atoms.
Based on the position of each atom in space relative to the window a weighting between 0 and 1 is thereby allocated.
The calculated density at a given position is then given by the average of the weighted atomic masses for the different atomic species.

\begin{figure}
    \includegraphics[width=3.4 in]{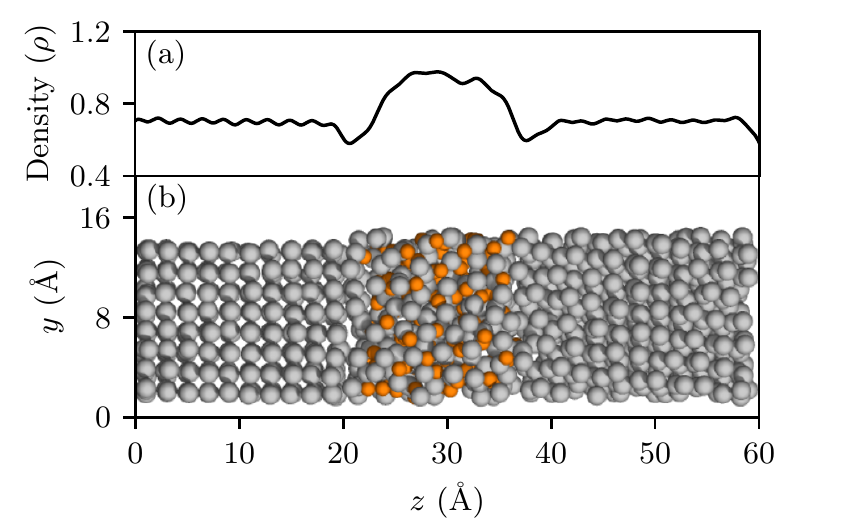}
    \caption{
        (a) The material density in the junction model as a function of position. 
        (b) The final atomic structure of a \alaloxal\ junction model where oxygen and aluminium atoms are depicted as orange and gray spheres respectively.
        This calculation was performed with the S-M potential.
    }
    \label{fig:junction}
\end{figure}

Figure~\ref{fig:junction}(a) shows an example of a density profile calculated in this way for the junction depicted in Fig.~\ref{fig:junction}(b).
The aluminium contacts have a density of approximately $\rho = 0.7$ with some oscillations due to the alignment of the lattice planes perpendicular to the $z$-axis.
There are notable drops in the density at the metal--oxide interfaces and the oxide region in the centre has a higher density than the contacts.

\begin{figure}
    \includegraphics[width=3.4 in]{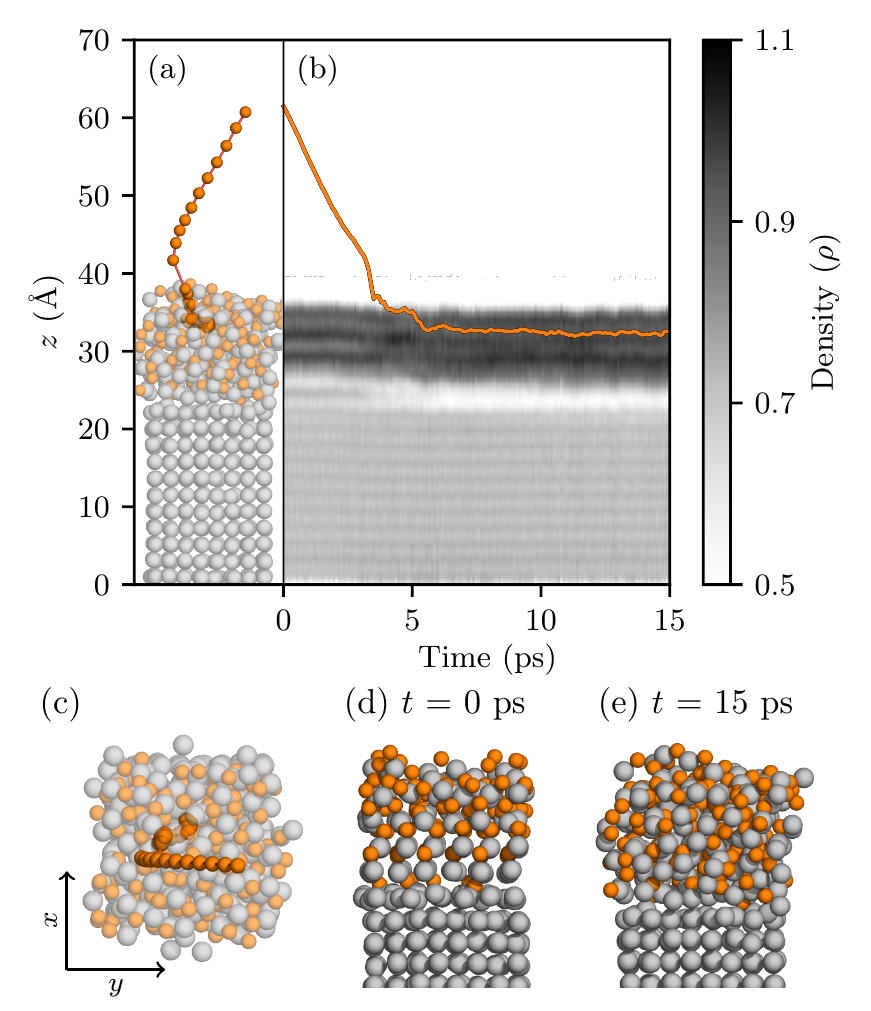}
    \caption{
        (a) The path of the deposited oxygen atom is overlaid on the final atomic positions for this iteration.
        (b) The density profile in the structure over the duration of the simulation.
        The position of the deposited oxygen atom is shown as an orange line.
        (c) The path of the oxygen atom from a perspective above the structure.
        (d) The structure at the beginning of this simulation at $t$ = 0~ps.
        (e) The structure at the end of this simulation at $t$ = 15~ps.
        This calculation was performed with the S-M potential.
    }
    \label{fig:phase_change}
\end{figure}

Figure~\ref{fig:phase_change}(a)--(c) show the path of the newly added oxygen atom as it approaches then bonds to the surface.
The evolution of the material density in the structure over the course of a 15~ps simulation is shown in Fig.~\ref{fig:phase_change}(b).
The spatially varying density at each time is calculated in the same way as for Fig.~\ref{fig:junction}(a).
We observe that the incoming oxygen atom -- which is embedded 4--5~\AA\ below the surface by the end of the simulation -- seems to initiate the transition from a semi-crystalline [Fig.~\ref{fig:phase_change}(d)] to an amorphous structure [Fig.~\ref{fig:phase_change}(e)] (see Supplemental Video 1).
This suggests that a locally ordered region of the growing oxide may undergo this type of phase change due to a single oxygen atom disrupting the structure, though we note that the details of this effect may be modified by the finite size of the simulation cell.
A more direct investigation of this transition may be possible using Monte Carlo based techniques.\cite{Zeng2016,Opletal2019}

\begin{figure}
    \includegraphics[width=3.4 in]{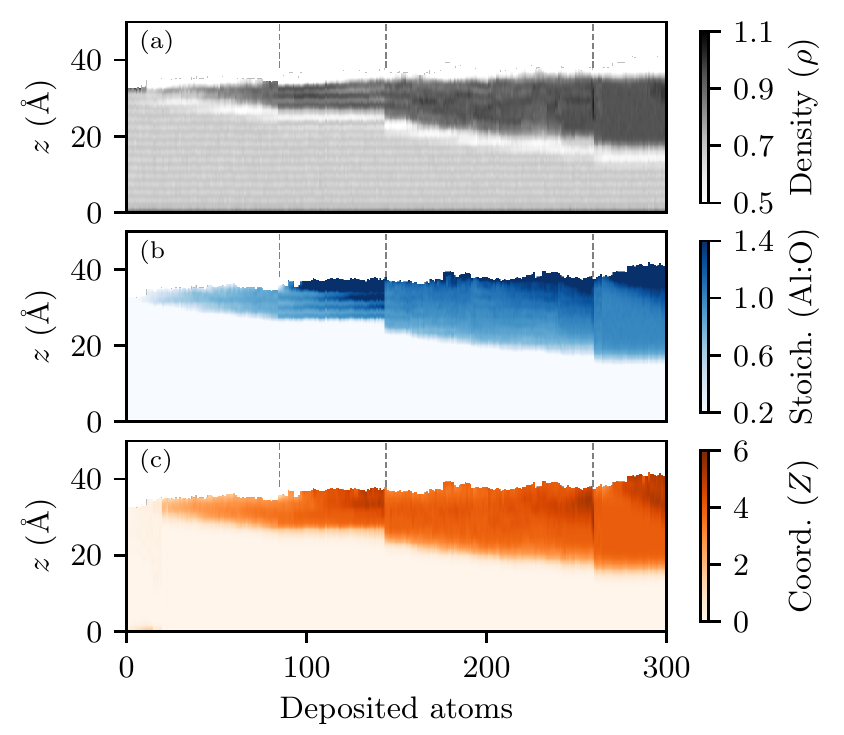}
    \caption{
        (a) Density of the aluminium--aluminium oxide structure shown as a function of position and deposition progress.
        The dashed lines mark the points where structural changes occur as the surface oxide changes form or a layer of the aluminium substrate is consumed by the oxidation process.
        (b) Variation of the stoichiometric ratio between oxygen and aluminium over the course of the oxidation simulation.
        (c) Coordination of oxygen atoms about aluminium over the course of the oxidation simulation.
        This calculation was performed with the S-M potential.
    }
    \label{fig:dynamic_oxidation_analysis}
\end{figure}

While Fig.~\ref{fig:phase_change} shows how the density changes over the course of a single 15~ps simulation, we can also examine how the structure evolves as the oxide growth is simulated.
In Fig.~\ref{fig:dynamic_oxidation_analysis}(a) the density is calculated as a function of $z$ at the end of each iteration, i.e. after a new oxygen has been added to the surface.
We use this to understand the evolution of the density profile as oxygen atoms are consecutively deposited on 16~$\times$~16~\AA$^2$ Al(100) surface.
There is a low density region at the lower Al/AlOx interface which is persistent and moves down in $z$ as more aluminium is incorporated into the growing oxide layer.
The points where significant structural changes occur -- such as the crystalline-to-amorphous transition depicted in Fig.~\ref{fig:phase_change} (the second dashed line at $N=144$ atoms) -- are marked with dashed lines.

In the same way we determine the density as a function of $z$ we calculate the spatial variation of the stoichiometry (O:Al ratio) and coordination number for each iteration [Fig.~\ref{fig:dynamic_oxidation_analysis}(b) and (c)].
The abrupt structural changes visible in the density data can also be seen in the stoichiometry and coordination.
Coordination numbers of aluminium atoms are calculated by counting the number of oxygen atoms within 2.4~\AA.
This distance corresponds to a position in the radial distribution function $g(r)$ between the first and second peaks.

In Fig.~\ref{fig:dynamic_oxidation_analysis}(b) we note that the AlOx/vacuum interface is oxygen rich compared with the Al/AlOx interface.
Figure~\ref{fig:dynamic_oxidation_analysis}(c) shows that more highly coordinated aluminium atoms tend to be towards the surface.
This is consistent with the analysis of the stoichiometry which shows that more oxygen atoms are available for bonding closer to the surface.

Oxidation calculations were performed on an Al(111) surface with the temperature of the thermostat set to correspond to experimental values of interest: liquid nitrogen cooled (77~K), room temperature (300~K), heated to 100$^{\circ}$C (370~K), and heated to 200$^{\circ}$C (470~K).
The temperature of the oxygen gas -- used to generate the velocities from a Maxwell--Boltzmann distribution -- was 300~K in all cases.
This is a computational approximation to experimental conditions where it is far more likely to be able to change the temperature of the substrate than the temperature of the gas introduced to the chamber.

\begin{figure}
    \includegraphics[width=3.4 in]{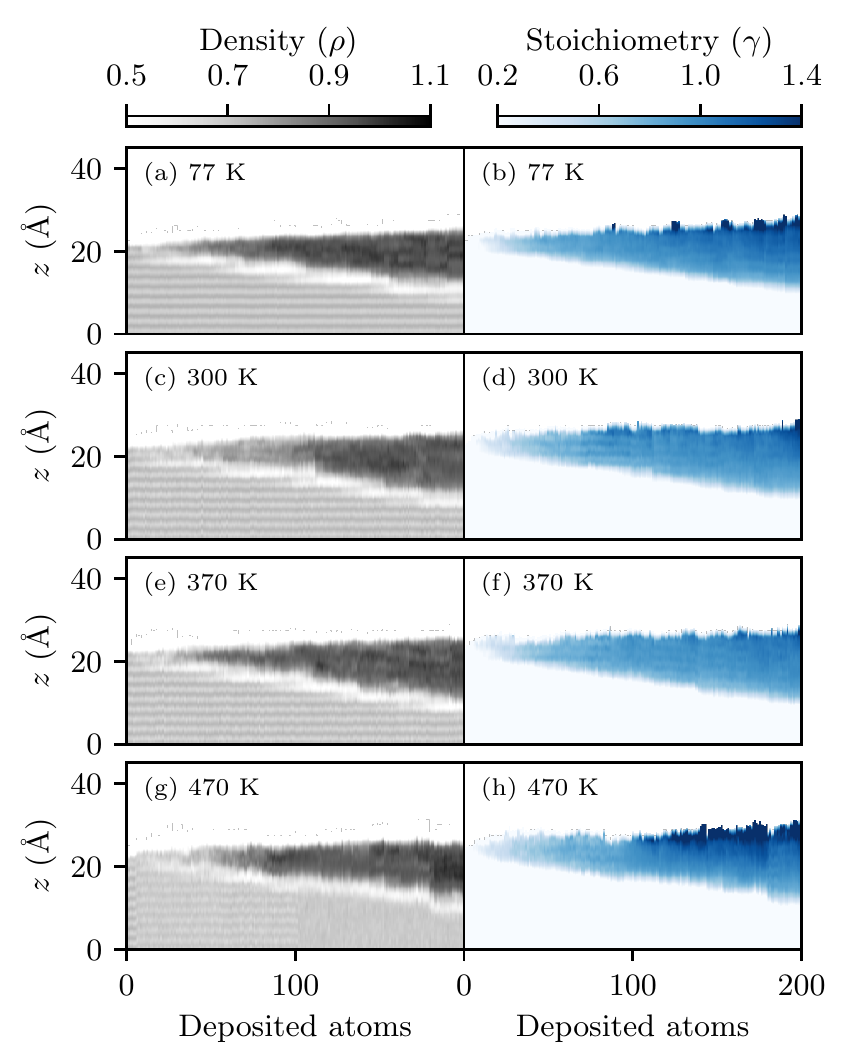}
    \caption{
        Development of the structure during simulated oxidation at 77, 300, 370, and 470 K.
        These calculations were performed with the S-M potential.
    }
    \label{fig:temperature_comparison}
\end{figure}

The evolution of the density and stoichiometry in the growing oxides is shown in Fig.~\ref{fig:temperature_comparison}.
Here we can again see the step-like way in which the aluminium substrate is converted into surface oxide. 
This is most clearly visible in the low temperature calculations.
The calculations proceed in general without the abrupt structural changes such as those in Fig.~\ref{fig:dynamic_oxidation_analysis} though some such features can be seen in panels (c) and (g).
There appears to be minimal difference between the 77~K and 370~K calculations other than the thermal atomic motion limiting the clarity of the calculated density.

\begin{figure}
    \includegraphics[width=3.4 in]{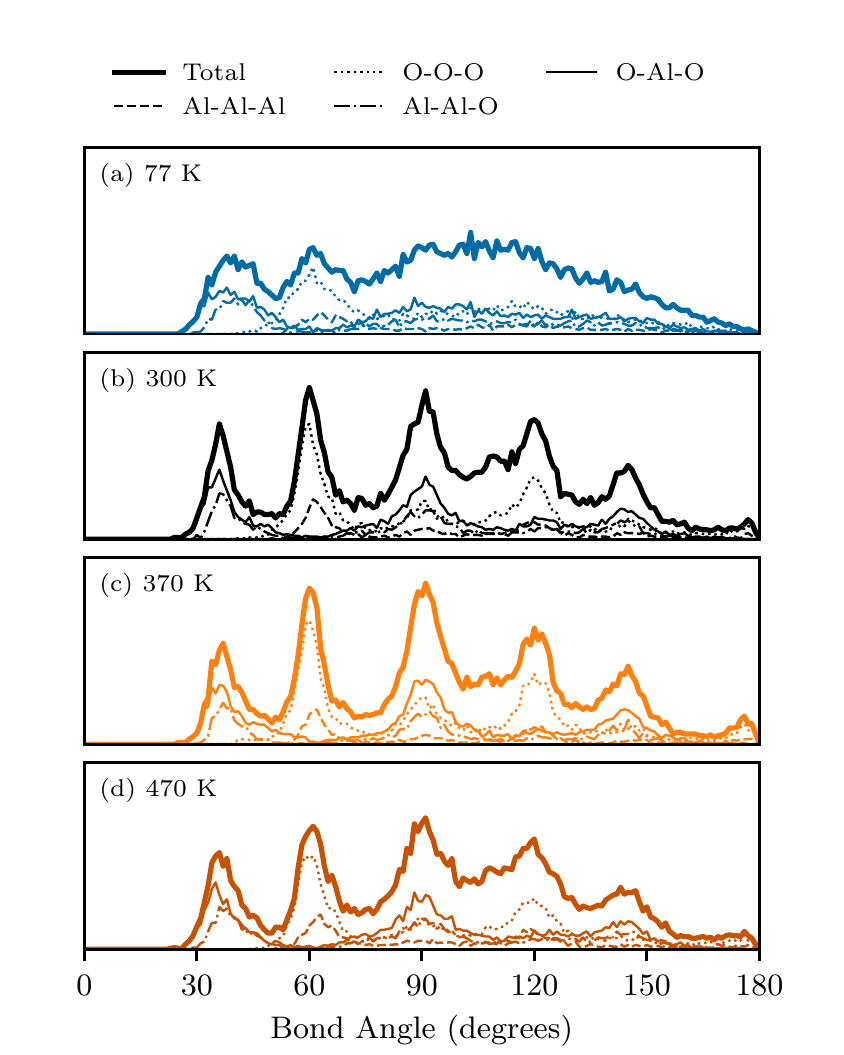}
    \caption{
        Bond angles in the surface oxide following simulated oxidation at 77, 300, 370, and 470 K.
        These calculations were performed with the S-M potential. 
    }
    \label{fig:bond_angles}
\end{figure}

By examining the bond angles in the oxide (Fig.~\ref{fig:bond_angles}) after 200 atoms have been deposited we can see a structural difference which is not evident in the density or stoichiometry.
The bond angle analysis shows strong peaks for temperatures of 300~K and 370~K indicating the presence of semi-crystalline structures in the oxide.
The high temperature calculation (470~K) has the same crystalline peaks which have been broadened slightly by the thermal noise.
By comparison the low temperature calculation (77~K) is significantly more amorphous.

\begin{figure}
    \includegraphics[width=3.4 in]{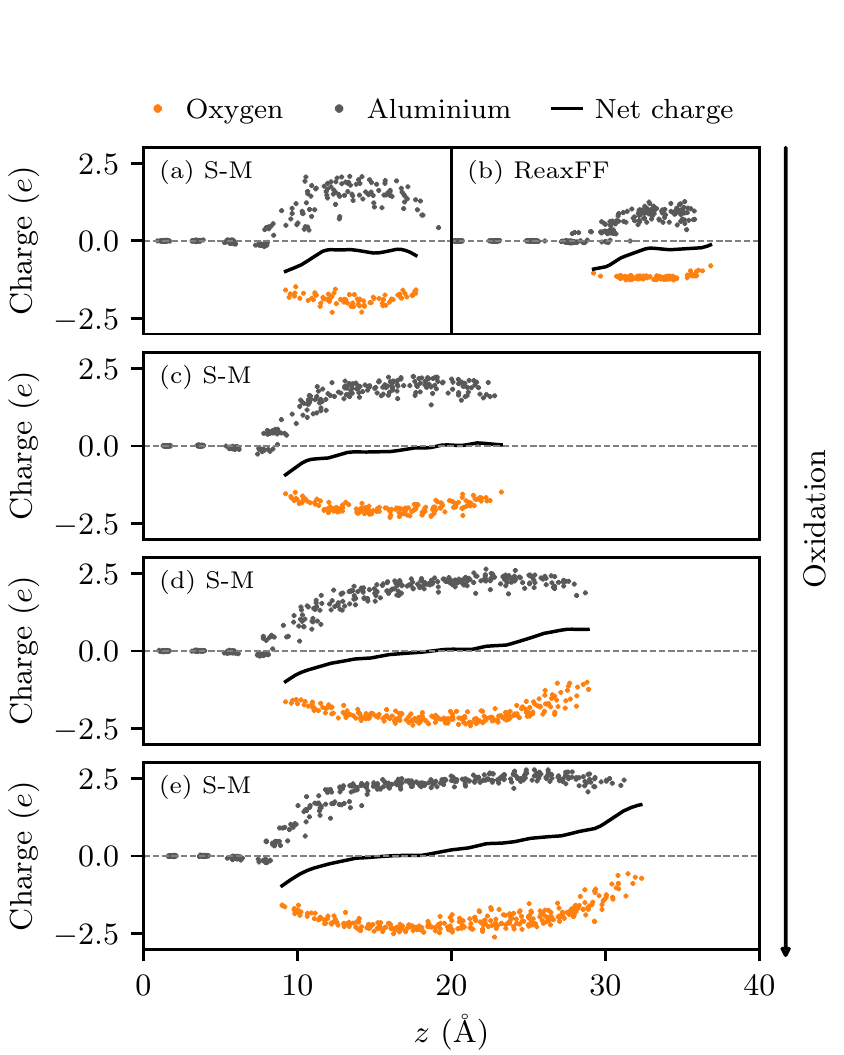}
    \caption{
        The partial charges of oxygen and aluminium atoms in the structure at varying stages of oxidation (shown as orange and gray dots respectively).
        The black lines show the net charge in the structure as a function of $z$.
        (a)--(b) A comparison between the S-M and ReaxFF potentials for a 1~nm thick surface oxide.
        (c)--e) The development of a charge gradient after continued oxidation using the S-M potential.
    }
    \label{fig:charges_oxide}
\end{figure}

Figure~\ref{fig:charges_oxide} shows how the distribution of charge in the system changes at different stages of oxide growth at 300~K.
The continued oxidation of the surface with the S-M potential gives rise to a charge gradient across the oxide.
This is in agreement with the empirical understanding of oxidation.
Mott-Cabrera oxidation theory is predicated on the effect of such a charge gradient on incoming oxygen atoms and molecules.\cite{Cabrera1948}
We also examine how the charge distribution differs between the two empirical potentials we have used [compare Fig.~\ref{fig:charges_oxide}(a) and (b)].
In both cases the net charge is neutral in the bulk of the oxide and tends to become negative at the metal--oxide interface, though the charge separation is smaller in magnitude by around a factor of two with the ReaxFF potential.
We are unable to compare the charges at the later stages of oxidation as the ReaxFF potential qualitatively reproduces the natural termination of the process at a limiting thickness (see following section).

\begin{figure}
    \includegraphics[width=3.4 in]{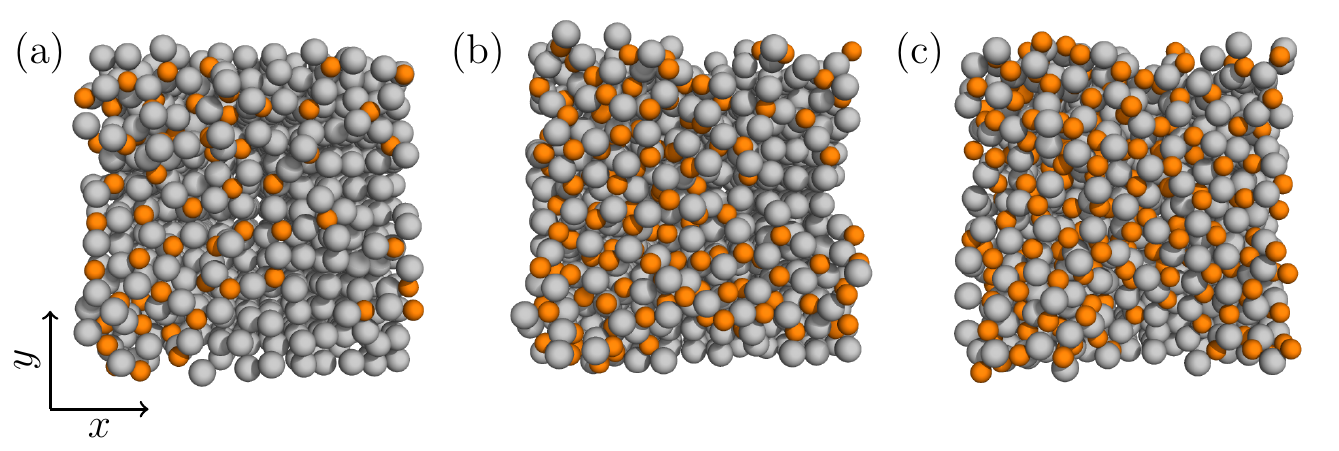}
    \caption{
        A hole is observed to form then close during the oxidation of a 24~$\times$~24~\AA$^2$ substrate.
        This calculation was performed with the S-M potential.
    }
    \label{fig:islands}
\end{figure}

The iterative method by which we form the oxide layer allows for a detailed study of the dynamics.
Figure~\ref{fig:islands} shows clustering of oxygen atoms on the aluminium surface forming a hole [evident in Fig.~\ref{fig:islands}(b)] which is filled in as the oxide continues to grow (see Supplemental Video 2).
Holes which form and close in this way have previously been observed in MD calculations.\cite{Zhou2007}
Experimentally, \citeauthor{Nguyen2018} observed the formation of islands at lattice shelves (terraces) on the aluminium surfaces by making a series of time-resolved observations of the growth of oxides on pristine Al(100) and Al(111) surface.\cite{Nguyen2018}
The islands proceed to grow laterally and merge to cover the remaining exposed aluminium. 

\subsection{Discussion}

The oxidation of aluminium is known to self-terminate when a thin amorphous oxide layer has been formed\cite{Jeurgens2002b,Cai2011} and, as the magnitude of the tunnelling current in Josephson junctions is exponentially dependent on the thickness of the oxide layer, the factors which affect the self-limiting thickness are important considerations for device design.\cite{Aref2014}
In order to optimise processes for device applications the effect on the uniformity and morphology of the barrier obtained by heating or cooling the aluminium crystal substrate, using single crystal substrates of different orientations, or varying the oxidation pressure has been studied.\cite{Nik2016,Fritz2019}

In our calculations with ReaxFF we observed self-limiting behaviour on both Al(100) and Al(111) surfaces (averaged over eight simulations of each crystal orientation) at thicknesses of:
\begin{itemize}
    \item Al(100): 7.87 $\pm$ 0.80~\AA
    \item Al(111): 8.54 $\pm$ 0.56~\AA
\end{itemize}
A calculation was determined to have reached the limiting stage when 25 oxygens atoms in a row had been reflected from the surface without bonding.
By comparison, limiting behaviour was never observed in the simulated oxidation of aluminium surfaces with the S-M potential.
New oxygen atoms continued to bond to the surface as other oxygen atoms are displaced deeper into the structure.
This proceeded until all of the aluminium atoms in the initial contact was incorporated into the growing aluminium oxide.

In recent studies the oxide thicknesses have been measured directly by taking images of the structure at nanometre scales with scanning transmission electron microscopy (STEM).\cite{Aref2014,Zeng2015a}
\citeauthor{Zeng2015a} measured the oxide thickness in this way at hundreds of positions for three \alaloxal\ samples.\cite{Zeng2015a}
Mean thicknesses of 1.66--1.88~nm are reported though the oxide thickness was measured to be as thin as 1.1~nm in some places and up to 2.2~nm thick in others.

As an alternative to measuring the barrier thickness in STEM images, an estimate of the average thickness over a large area can be made by comparing the relative intensities of the aluminium metal and aluminium oxide signals obtained from x-ray photoemission spectroscopy (XPS).\cite{Jeurgens1999}
Measurements of the limiting thickness made in this way find values in the range 5--10~\AA\ for Al(100) and Al(111) substrates.\cite{Reichel2008,Jeurgens2002c}
For Al(111) surfaces oxidised at room temperature over a wide range of pressures between $1 \times 10^{-6}$~Pa and 650~Pa the self-limiting thickness was found to increase monotonically from 0.2 to 1.2~nm.\cite{Cai2011}
Another similar study reports self-limiting thicknesses of 0.49--1.36~nm on Al(100) and Al(111) surfaces for partial pressures from $1 \times 10^{-5}$~Pa to 1.0~Pa.\cite{Flototto2015}
\citeauthor{Nguyen2018} find that Al(111) surfaces have slightly thicker oxides than Al(100) surfaces while the oxide thickness increases from 0.95~nm to 2.6~nm as the pressure changes from $4 \times 10^{-5}$ to $4 \times 10^{-3}$~Pa.\cite{Nguyen2018}

\citeauthor{Sankaranarayanan2009} simulate the oxidation of Al(100) with both O atoms and O$_2$ molecules by maintaining a particular number density of oxygen around a aluminium crystal structure.\cite{Sankaranarayanan2009}
A self-limiting oxide thickness of 1.6~nm is reported as well as low densities at the AlOx/Al interfaces.
Due to the manner in which we approximate the deposition process, our work is most reasonably compared to the lowest pressure experimental reports which also produce the thinnest oxide layers.
The thicknesses we report here are of the same order as existing experimental and computational reports although, based on the most recent studies, they are likely to be lower than the true values.

Many studies which report thickness also investigate the composition of the oxide layer (i.e. ratio of oxygen to aluminium) which can also be estimated from XPS measurements.\cite{Jeurgens1999}
On Al(100) and Al(111) substrates oxides have been reported to be super-stoichiometric with final O:Al ratios of 1.6--1.7.\cite{Cai2011,Flototto2015}
The stoichiometry at the surface has been found to be lower than that in the centre of the oxide.\cite{Nguyen2018}
The overall composition of the oxide on Al(431) substrates is reported to be stoichiometric (O:Al~=~1.5) whereas the surface is highly substoichiometric (O:Al~=~0.3--0.7).\cite{Jeurgens2002b}

\citeauthor{Fritz2018} report stoichiometries for oxides grown using four different techniques: thermal oxidation with and without UV illumination, plasma oxidation, and physical vapor deposition achieved by heating Al$_2$O$_3$-pellets with an electron beam.\cite{Fritz2018}
The stoichiometries were determined using STEM electron energy loss spectroscopy and are in the range 1.1--1.3 in the amorphous oxide regions except for the thermally oxidised sample without UV illumination which has a reported stoichimetry of 0.5.
In some cases nanocrystals of either Al or stoichiometric $\gamma$-\corundum\ are formed.

High oxygen concentrations at the surface, such as those we report, are in agreement with other computational work on the topic.
\citeauthor{Zeng2016} investigated the microstructure of an oxide barrier with STEM imaging.\cite{Zeng2016}
Based on these measurements an atomistic model of a possible tunnel barrier structure is then reconstructed which predicts oxygen deficiency at the Al/AlOx interfaces.
\citeauthor{Sankaranarayanan2009} also find higher oxygen concentrations at the AlOx/gas interface than the AlOx/Al interface in their simulations.\cite{Sankaranarayanan2009}

\citeauthor{Jeurgens2002a} observe a change from amorphous to crystalline morphology in oxide layers at the temperature was raised from 573 to 773~K.\cite{Jeurgens2002a}
A change from amorphous to semi-crystalline ``$\gamma$(-like)-\corundum'' structures was observed at between 400 and 550~K by \citeauthor{Reichel2008a} depending on the crystallographic orientation of the lower Al substrate.\cite{Reichel2008a}
We observe crystalline features in the bond angle distribution even at room temperature (300~K).
These features are not present at 77~K which suggests that the temperature at which the amorphous--crystalline transition takes place may be reduced by the periodic boundary conditions in the simulation.
A more detailed future study focusing solely on temperature effects and using a range of substrate sizes would be ideal to further understand this effect.

\section{Junction formation}
\label{sec:junction}

\subsection{Methodology}

Relatively few attempts have been made to construct complete \textsl{ab-initio} junction models, and those that exist are mostly limited by the high computational cost of density functional theory (DFT) calculations.
These models have been created by placing a stoichiometric layer of \corundum\ between two metallic contacts of either pure aluminium or niobium and do not include any disorder in the oxide layer.\cite{ZemanovaDieskova2013,Jung2009}
Junction models developed using a simulated annealing method provide a more accurate representation of the real oxide layer which is known to be amorphous.\cite{DuBois2016}
We have recently reported transport properties of junction models formed with simulated annealing.\cite{Cyster2020}

When working with the S-M potential, rather than continuing the oxidation indefinitely, we create junction models by beginning to deposit aluminium on the surface when the oxide reaches the desired thickness ($\sim$1.4--1.6~nm).
The oxides grown with ReaxFF self-limit at a given thickness after which we start depositing aluminium.
The methodology for aluminium deposition is the same as for the oxidation, excepting that the velocities are selected from a normal distribution with a mean of approximately 600~m/s and a standard deviation of 20~m/s.
These values are representative of the evaporation method of thin-film deposition which is used experimentally.\cite{Curran1982}
The second aluminium electrode is grown until it is of a similar thickness to the initial aluminium contact region.

\subsection{Results}

\begin{figure}[b]
    \includegraphics[width=3.4 in]{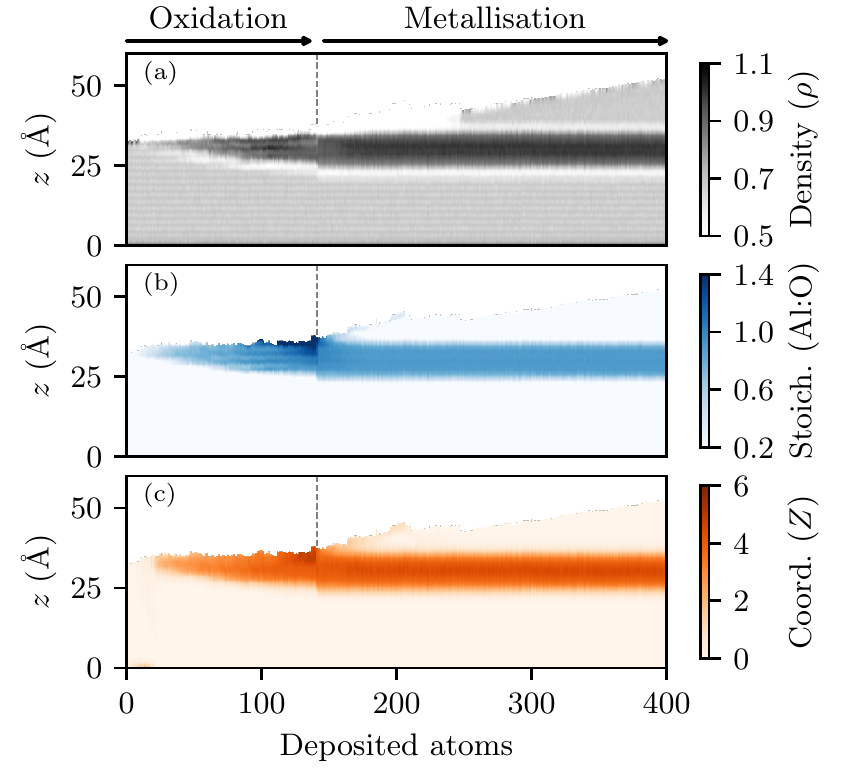}
    \caption{
        Evolution of the (a) density, (b) stoichiometry, and (c) coordination as the growth of a complete \alaloxal\ junction is simulated.
        The change from oxygen to aluminium deposition is indicated by the vertical dashed lines.
        This calculation was performed with the S-M potential.
    }
    \label{fig:dynamic_junction_analysis}
\end{figure}

Figure~\ref{fig:dynamic_junction_analysis} shows how the density, stoichiometry and coordination evolve during the creation of a deposited junction.
Oxygen atoms are consecutively added to a 16~$\times$~16~\AA$^2$ Al(100) surface until the oxide layer reaches a thickness of 1.4~nm.
After the oxide layer is formed, aluminium is deposited to form the second electrode of the junction structure.
The vertical dashed lines indicate the point at which this change from oxygen to aluminium deposition takes place.

The development of low density regions at the AlOx/Al interfaces is visible in Fig.~\ref{fig:dynamic_junction_analysis}(a).
In Fig.~\ref{fig:dynamic_junction_analysis}(b) we observe an oxygen rich surface at the end of the oxidation process in agreement with Fig.~\ref{fig:dynamic_oxidation_analysis}.
New aluminium atoms quickly bond to this surface oxygen and the stoichiometries at both AlOx/Al interfaces become equivalent (see Supplemental Video 3).

\begin{figure}
    \includegraphics[width=3.4 in]{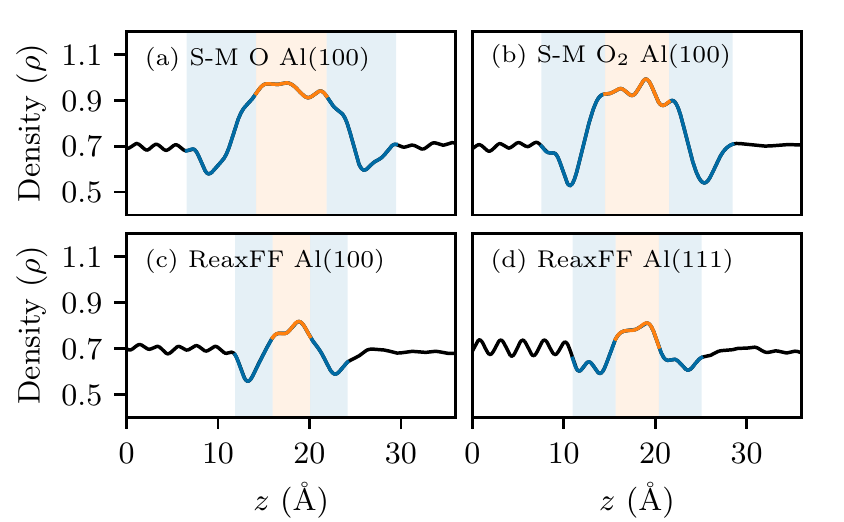}
    \caption{
        (a)--(d) The density of Al/AlOx/Al junction models as a function of $z$ (averaged over the $x$ and $y$ dimensions).
        Low densities are consistently observed at the \alalox\ interfaces.
        The blue and orange shaded regions are used to determine the minimum and mean densities respectively for Fig.~\ref{fig:density_histograms}.
        This calculation was performed with the S-M potential.
    }
    \label{fig:density_profiles}
\end{figure}

The spatial variation in the material density for four finished junction models is shown in Fig.~\ref{fig:density_profiles}.
The structure in panel b was formed by oxidising the surface with O$_2$ molecules rather than single O atoms and showed no discernible difference in any of our analyses.
Low densities are again observed at the interfaces between the contacts and the oxide.
This is a common feature in our analysis of the density as a function of position regardless of the crystal orientation, temperature of the aluminium substrate, or the empirical potential used.
The interfacial and central regions of the structure are shaded in blue and orange respectively.
The bounds of these regions are found by first taking the positions of the outermost oxygen atoms $O_L$ and $O_R$ to determine the thickness of the oxide layer $d = O_R - O_L$.
We then use four points along the $z$-axis $(O_L - d/4,\ O_L+d/4,\ O_R-d/4,\ O_L+d/4)$ to determine the boundaries of three regions with width $d / 2$: one in the centre of the oxide and one bridging each interface.
Considering the regions in Fig.~\ref{fig:density_profiles} by eye this appears to be a good heuristic approach to defining the central and interfacial regions.

\begin{figure}
    \includegraphics[width=3.4 in]{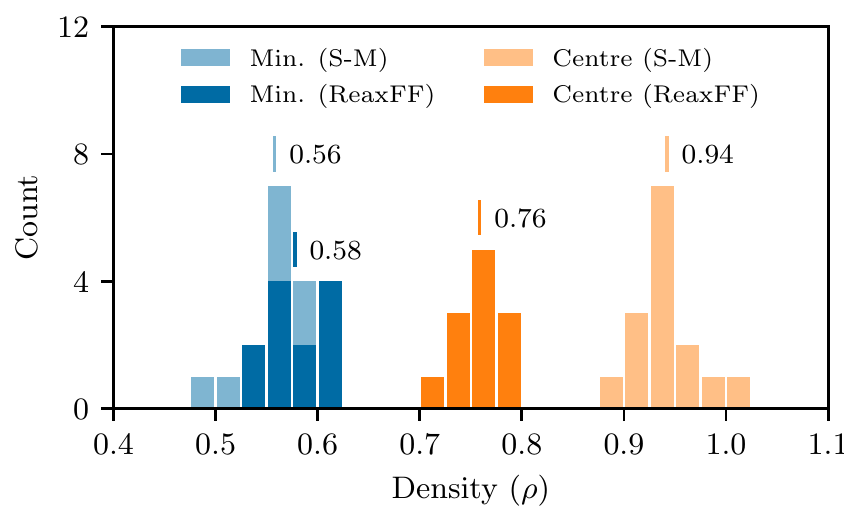}
    \caption{
        The distribution of minimum and mean densities in junction models deposited on 16~$\times$~16~\AA$^2$ Al(100) substrates with different potentials.
        The minimum density is the minimum value in the blue shaded regions in Fig.~\ref{fig:density_profiles}.
        The central density is the mean value of the orange shaded region.
        As shown in Fig.~\ref{fig:density_profiles} the minimum densities occur at the Al/AlOx interfaces.
    }
    \label{fig:density_histograms}
\end{figure}

Fig.~\ref{fig:density_histograms} shows histograms of the minimum and central density for a range of junction models formed with both the S-M and ReaxFF potentials.
The minimum densities are determined from the lowest density value in the blue shaded interfacial regions in Fig.~\ref{fig:density_profiles} and the centre density is the mean of the orange shaded region at the centre.
Both potentials predict a reduced density at the interface ($\rho$~=~0.56--0.58) which is a persistent feature across all simulations.
Junctions deposited with the S-M potential have a higher density in the centre of the oxide.

\begin{figure}
    \includegraphics[width=3.4 in]{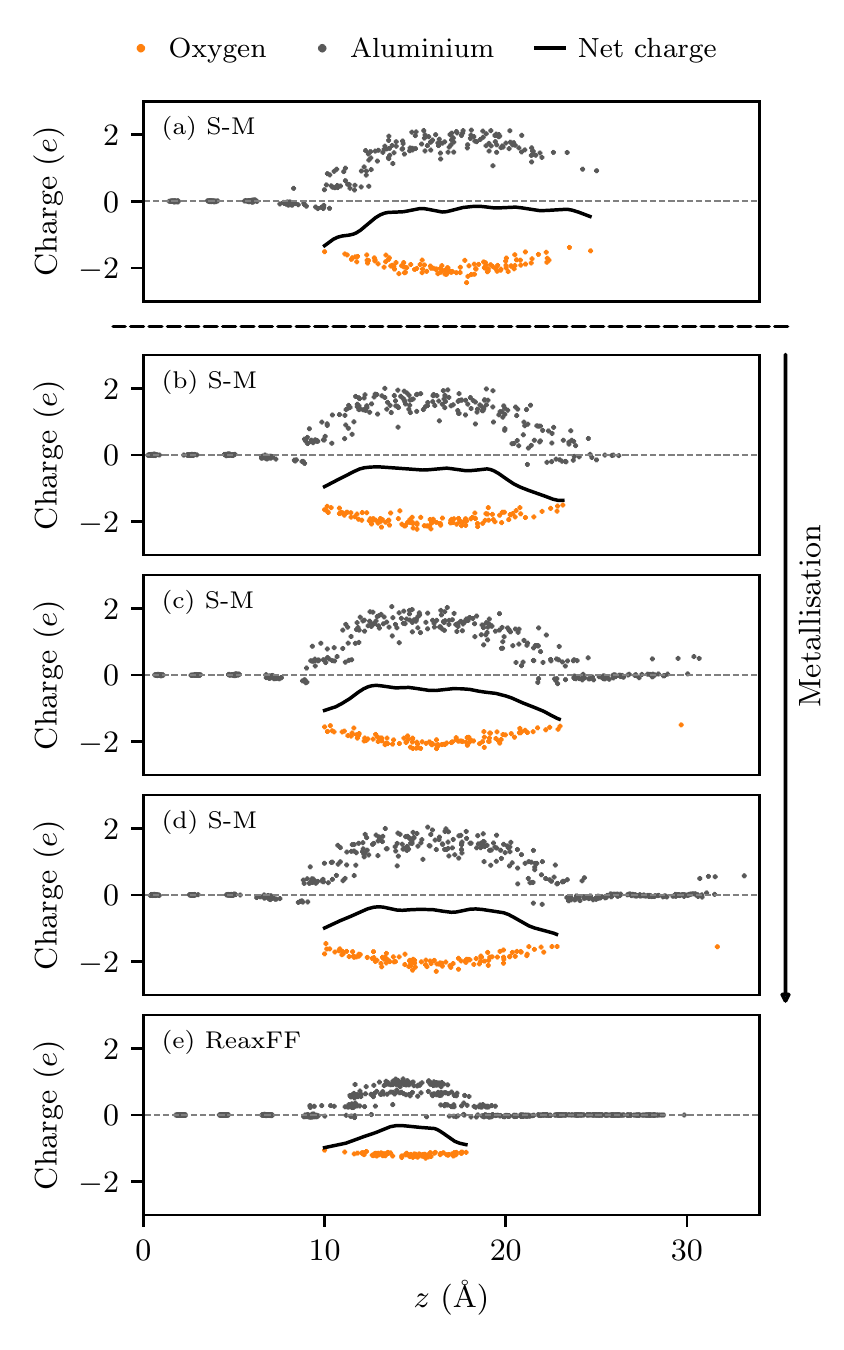}
    \caption{
        The partial charges of atoms in the structure as aluminium atoms are added to the finished \alalox surface.
        The black lines show the net charge in the structure is as a function of $z$.
        (a) The initial oxidised aluminium surface.
        (b)--(d) Continued metallisation with the S- potential.
        (e) Partial charges in the final junction structure formed with the ReaxFF potential.
    }
    \label{fig:charges_metal}
\end{figure}

Looking at the partial charges on the atoms as we add aluminium to the oxide surface (Fig.~\ref{fig:charges_metal}(a)--(d) we can see the shape of the net charge become negative at the interfaces and neutral in the centre of the barrier.
The same profile is observed for a junction created with ReaxFF (Fig.~\ref{fig:charges_metal}e) though the barrier is significantly thinner due to the self-limiting of the oxidation calculation.
As in Fig.~\ref{fig:charges_oxide} we observe that the magnitude of the charge separation is reduced relative to the S-M results.

\begin{figure}
    \includegraphics[width=3.4 in]{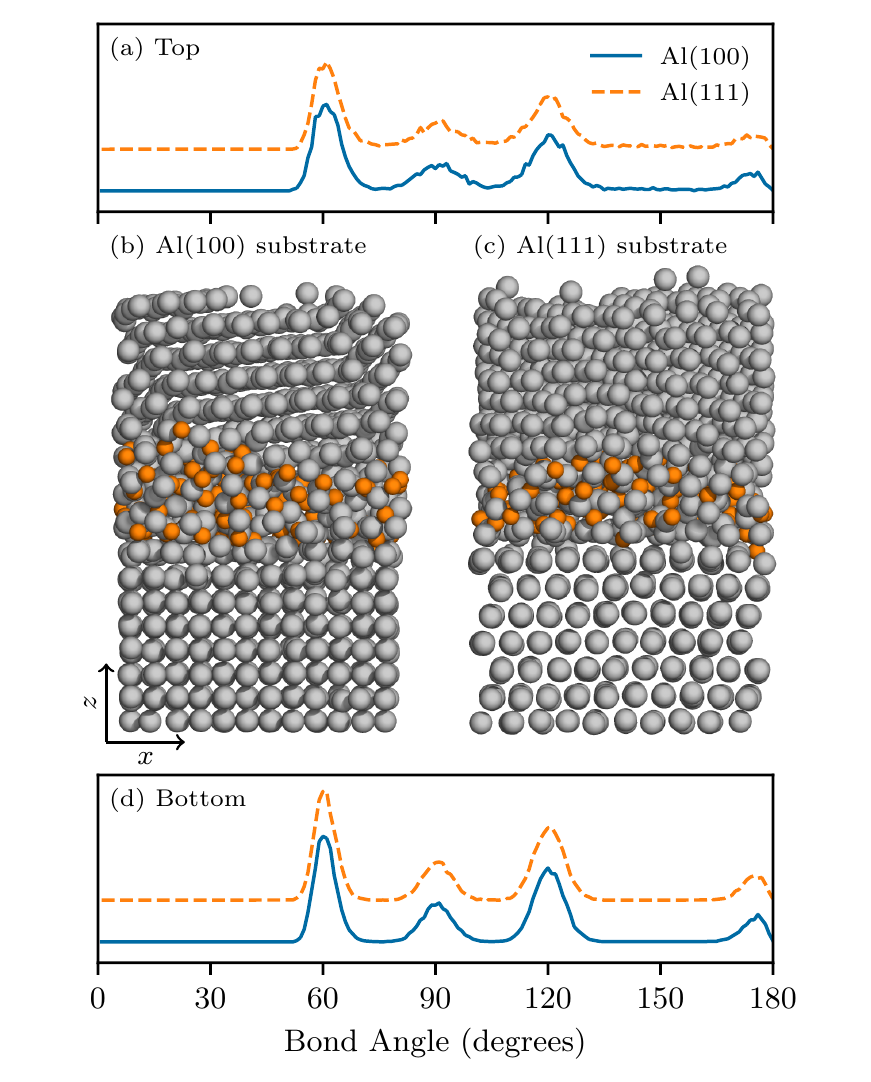}
    \caption{
        Bond angles in the top (a) and bottom (d) aluminium contacts on 24~$\times$~24~\AA$^2$ substrates.
        The junction structures grown on Al(100) and Al(111) substrates are shown in panels (b) and (c) respectively.
        This calculation was performed with the ReaxFF potential.
    }
    \label{fig:bond_angles_al}
\end{figure}

We use the bond angles in the aluminium contacts as a measure of the crystallinity of the structures.
Fig.~\ref{fig:bond_angles_al}(d) shows the bond angles calculated for the Al(100)
and Al(111) substrates we generate at the beginning of the oxidation simulation after thermalisation at 300~K.
The data shown in Fig.~\ref{fig:bond_angles_al}(a) is for the deposited aluminium contacts demonstrating that the crystal structure forms naturally as a result of the atomic interactions described by the empirical potential.
Figure~\ref{fig:bond_angles_al}(b) and (c) show the junction structures as grown on Al(100) and Al(111) substrates respectively.
The ordering of the atomic layers is partially evident in these images however it is somewhat obscured as the orientation of the top contact is not aligned with the substrate direction.

\subsection{Discussion}

The apparent density in the centre of the oxide increases as it grows. 
This effect can be seen in Fig.~\ref{fig:dynamic_oxidation_analysis}(a), \ref{fig:temperature_comparison}, and \ref{fig:dynamic_junction_analysis}(a).
From Fig.~\ref{fig:density_histograms} we see that the two empirical potentials give similar densities at the metal-oxide interface while S-M predicts a higher density at the centre of the oxide than ReaxFF, although part of this discrepancy may be caused by the reduced oxide thickness in the ReaxFF junctions.
The density of \alox\ barriers formed with thermal oxidation is not widely reported in the literature (possibly due to the difficulty of measuring the nm-thick layer).
Studies which use different experimental methods\cite{Sullivan1998a,Koski1999} to deposit thicker layers (~1$\mu$m) report densities in a wide range from $\rho$ = 0.58 to 0.95.
Oxide densities reported in simulations of thin film oxides\cite{Gutierrez2000,Campbell2005,Colleoni2015,Zeng2016,Trybula2019} lie within a narrower range of $\rho$ = 0.73--0.88.

Spatial variation of the density is evident in many of our results with a pronounced reduction at the metal-oxide interface.
This is in agreement with Auger analysis by \citeauthor{Evangelisti2017} which ``suggests density variations across the oxide layer, with lower densities near the surface and the metal-oxide interface.''\cite{Evangelisti2017}
The authors also note that they measured minimal variation in the stoichiometry across the thickness of the oxide which is in agreement with our results in Fig.~\ref{fig:dynamic_junction_analysis}(b).

\citeauthor{Fritz2019} achieved epitaxial growth of an Al(111) layer on a clean Si(111) substrate.\cite{Fritz2019}
In this case thickness fluctuations in the \alox\ are minimised and matching of the crystallographic orientation between the lower and upper aluminium layers is observed.
In the present work we observe crystallinity in both aluminium layers but no alignment between the top and bottom contacts.
It would be an interesting extension to perform junction formation calculations as a function of temperature and the thickness of the oxide layer to increase our understanding of how this information is transferred across the oxide layer.

\section{Conclusions}

Using our novel iterative approach to oxide growth we have created \alaloxal\ junction models with both the S-M and the ReaxFF potentials.
A key difference in the behaviour of the potentials is that ReaxFF qualitatively reproduces the self-limiting behaviour which is observed experimentally.
The final densities of the oxides formed with ReaxFF are closer to the mean of the experimental reports though the densities in the S-M models are still within the experimental range.
Without more accurate reports of the oxide density for direct comparison it is difficult to comment on the reliability of the empirical potential in faithfully reproducing the physics of the oxide formation.
Making a comparison in relative terms, ReaxFF is a more modern potential which qualitatively reproduces results closer to experimental reports.
It is possible that a reparameterisation of the force field for the oxidation of aluminium surfaces rather than nanoclusters may further improve the accuracy of the results.

In general, \textsl{ab-initio} models of \alaloxal\ junctions are difficult to develop due to the inherently amorphous oxide layer.
The iterative approach we adopt in building the oxide layer atom by atom allows us to see dynamic changes in the structure that would be missed when creating oxide models with simulated annealing.
The formation and closing of holes in the oxide, the transition of surface oxide between amorphous and semi-crystalline configurations, and the development of a charge gradient are all examples of these observations.
We believe this type of simulation to be a promising approach as many results in the present work -- such as self-limiting oxidation, the trend of temperature dependence of the oxide crystallinity, the reduced density at \alalox\ interfaces, and the crystallisation of the deposited aluminium contacts -- are in line with experimental reports.

We also note that the iterative deposition approach is easily adaptable to study other thin film deposition processes, provided that an empirical potential is used which appropriately describes the interactions between the different atomic species.
For example, experimental evidence of an amorphous interface layer consisting of Al, Si and O between the bottom aluminium contact and the silicon substrate has been reported.\cite{Zeng2015}
It may be possible to observe the development of this interface layer by including the silicon substrate in the simulation and performing an iterative oxidation calculation.

The growth of ultra-thin oxide layers is relevant to the manufacturing of many different devices.
Single-barrier junctions which use superconductors such an aluminium or niobium can be used as Josephson junctions.\cite{Kohlstedt1993,Wu2013}
Double-barrier junctions constructed with aluminium and aluminium oxides are used in magnetic tunnel junctions (MTJs).\cite{Zhu2006}
Other materials are often used in MTJs such as in CoFeB--MgO--CoFeB junctions.\cite{Zhu2006}
While some concepts for creating magneto-resistive random access memory (MRAM) have even more exotic geometries, all of these devices make use of at least one thin oxide layer in their design.\cite{Zhu2006}

\begin{acknowledgments}
    The authors acknowledge support of the Australian Research Council through grants DP140100375, CE170100026 (MJC), and CE170100039 (JSS).
    The authors also acknowledge useful discussions with P.~Delsing, S.~Fritz, J.~Gale, D.~Gerthsen, N.~Katz, E.~Olsson, J.~Pekola, and L.~Zeng.
    This research was undertaken with the assistance of resources from the National Computational Infrastructure (NCI), which is supported by the Australian Government. 
    The authors acknowledge the people of the Woi wurrung and Boon wurrung language groups of the eastern Kulin Nations on whose unceded lands we work.
    We also acknowledge the Ngunnawal people, the Traditional Custodians of the Australian Capital Territory where NCI is located.
    We respectfully acknowledge the Traditional Custodians of the lands and waters across Australia and their Elders: past, present, and emerging.
\end{acknowledgments}

\bibliography{deposition}

\begin{thebibliography}{73}%
\makeatletter
\providecommand \@ifxundefined [1]{%
 \@ifx{#1\undefined}
}%
\providecommand \@ifnum [1]{%
 \ifnum #1\expandafter \@firstoftwo
 \else \expandafter \@secondoftwo
 \fi
}%
\providecommand \@ifx [1]{%
 \ifx #1\expandafter \@firstoftwo
 \else \expandafter \@secondoftwo
 \fi
}%
\providecommand \natexlab [1]{#1}%
\providecommand \enquote  [1]{``#1''}%
\providecommand \bibnamefont  [1]{#1}%
\providecommand \bibfnamefont [1]{#1}%
\providecommand \citenamefont [1]{#1}%
\providecommand \href@noop [0]{\@secondoftwo}%
\providecommand \href [0]{\begingroup \@sanitize@url \@href}%
\providecommand \@href[1]{\@@startlink{#1}\@@href}%
\providecommand \@@href[1]{\endgroup#1\@@endlink}%
\providecommand \@sanitize@url [0]{\catcode `\\12\catcode `\$12\catcode
  `\&12\catcode `\#12\catcode `\^12\catcode `\_12\catcode `\%12\relax}%
\providecommand \@@startlink[1]{}%
\providecommand \@@endlink[0]{}%
\providecommand \url  [0]{\begingroup\@sanitize@url \@url }%
\providecommand \@url [1]{\endgroup\@href {#1}{\urlprefix }}%
\providecommand \urlprefix  [0]{URL }%
\providecommand \Eprint [0]{\href }%
\providecommand \doibase [0]{http://dx.doi.org/}%
\providecommand \selectlanguage [0]{\@gobble}%
\providecommand \bibinfo  [0]{\@secondoftwo}%
\providecommand \bibfield  [0]{\@secondoftwo}%
\providecommand \translation [1]{[#1]}%
\providecommand \BibitemOpen [0]{}%
\providecommand \bibitemStop [0]{}%
\providecommand \bibitemNoStop [0]{.\EOS\space}%
\providecommand \EOS [0]{\spacefactor3000\relax}%
\providecommand \BibitemShut  [1]{\csname bibitem#1\endcsname}%
\let\auto@bib@innerbib\@empty
%</preamble>
\bibitem [{\citenamefont {Makhlin}\ \emph {et~al.}(2001)\citenamefont
  {Makhlin}, \citenamefont {Sch{\"{o}}n},\ and\ \citenamefont
  {Shnirman}}]{Makhlin2001}%
  \BibitemOpen
  \bibfield  {author} {\bibinfo {author} {\bibfnamefont {Y.}~\bibnamefont
  {Makhlin}}, \bibinfo {author} {\bibfnamefont {G.}~\bibnamefont
  {Sch{\"{o}}n}}, \ and\ \bibinfo {author} {\bibfnamefont {A.}~\bibnamefont
  {Shnirman}},\ }\href {\doibase 10.1103/RevModPhys.73.357} {\bibfield
  {journal} {\bibinfo  {journal} {Reviews of Modern Physics}\ }\textbf
  {\bibinfo {volume} {73}},\ \bibinfo {pages} {357} (\bibinfo {year} {2001})},\
  \Eprint {http://arxiv.org/abs/0011269} {arXiv:0011269 [cond-mat]}
  \BibitemShut {NoStop}%
\bibitem [{\citenamefont {Wendin}\ and\ \citenamefont
  {Shumeiko}(2007)}]{Wendin2007}%
  \BibitemOpen
  \bibfield  {author} {\bibinfo {author} {\bibfnamefont {G.}~\bibnamefont
  {Wendin}}\ and\ \bibinfo {author} {\bibfnamefont {V.~S.}\ \bibnamefont
  {Shumeiko}},\ }\href {\doibase 10.1063/1.2780165} {\bibfield  {journal}
  {\bibinfo  {journal} {Low Temperature Physics}\ }\textbf {\bibinfo {volume}
  {33}},\ \bibinfo {pages} {724} (\bibinfo {year} {2007})}\BibitemShut
  {NoStop}%
\bibitem [{\citenamefont {Clarke}\ and\ \citenamefont
  {Wilhelm}(2008)}]{Clarke2008}%
  \BibitemOpen
  \bibfield  {author} {\bibinfo {author} {\bibfnamefont {J.}~\bibnamefont
  {Clarke}}\ and\ \bibinfo {author} {\bibfnamefont {F.~K.}\ \bibnamefont
  {Wilhelm}},\ }\href {\doibase 10.1038/nature07128} {\bibfield  {journal}
  {\bibinfo  {journal} {Nature}\ }\textbf {\bibinfo {volume} {453}},\ \bibinfo
  {pages} {1031} (\bibinfo {year} {2008})}\BibitemShut {NoStop}%
\bibitem [{\citenamefont {Martinis}(2009)}]{Martinis2009a}%
  \BibitemOpen
  \bibfield  {author} {\bibinfo {author} {\bibfnamefont {J.~M.}\ \bibnamefont
  {Martinis}},\ }\href {\doibase 10.1007/s11128-009-0105-1} {\bibfield
  {journal} {\bibinfo  {journal} {Quantum Information Processing}\ }\textbf
  {\bibinfo {volume} {8}},\ \bibinfo {pages} {81} (\bibinfo {year}
  {2009})}\BibitemShut {NoStop}%
\bibitem [{\citenamefont {Wu}\ \emph {et~al.}(2013)\citenamefont {Wu},
  \citenamefont {Deng}, \citenamefont {Yu}, \citenamefont {Xue}, \citenamefont
  {Tian}, \citenamefont {Li}, \citenamefont {Chen}, \citenamefont {Zhao},\ and\
  \citenamefont {Zheng}}]{Wu2013}%
  \BibitemOpen
  \bibfield  {author} {\bibinfo {author} {\bibfnamefont {Y.-L.}\ \bibnamefont
  {Wu}}, \bibinfo {author} {\bibfnamefont {H.}~\bibnamefont {Deng}}, \bibinfo
  {author} {\bibfnamefont {H.-F.}\ \bibnamefont {Yu}}, \bibinfo {author}
  {\bibfnamefont {G.-M.}\ \bibnamefont {Xue}}, \bibinfo {author} {\bibfnamefont
  {Y.}~\bibnamefont {Tian}}, \bibinfo {author} {\bibfnamefont {J.}~\bibnamefont
  {Li}}, \bibinfo {author} {\bibfnamefont {Y.-F.}\ \bibnamefont {Chen}},
  \bibinfo {author} {\bibfnamefont {S.-P.}\ \bibnamefont {Zhao}}, \ and\
  \bibinfo {author} {\bibfnamefont {D.-N.}\ \bibnamefont {Zheng}},\ }\href
  {\doibase 10.1088/1674-1056/22/6/060309} {\bibfield  {journal} {\bibinfo
  {journal} {Chinese Physics B}\ }\textbf {\bibinfo {volume} {22}},\ \bibinfo
  {pages} {060309} (\bibinfo {year} {2013})}\BibitemShut {NoStop}%
\bibitem [{\citenamefont {Wang}\ \emph {et~al.}(2015)\citenamefont {Wang},
  \citenamefont {Axline}, \citenamefont {Gao}, \citenamefont {Brecht},
  \citenamefont {Chu}, \citenamefont {Frunzio}, \citenamefont {Devoret},\ and\
  \citenamefont {Schoelkopf}}]{Wang2015}%
  \BibitemOpen
  \bibfield  {author} {\bibinfo {author} {\bibfnamefont {C.}~\bibnamefont
  {Wang}}, \bibinfo {author} {\bibfnamefont {C.}~\bibnamefont {Axline}},
  \bibinfo {author} {\bibfnamefont {Y.~Y.}\ \bibnamefont {Gao}}, \bibinfo
  {author} {\bibfnamefont {T.}~\bibnamefont {Brecht}}, \bibinfo {author}
  {\bibfnamefont {Y.}~\bibnamefont {Chu}}, \bibinfo {author} {\bibfnamefont
  {L.}~\bibnamefont {Frunzio}}, \bibinfo {author} {\bibfnamefont {M.~H.}\
  \bibnamefont {Devoret}}, \ and\ \bibinfo {author} {\bibfnamefont {R.~J.}\
  \bibnamefont {Schoelkopf}},\ }\href {\doibase 10.1063/1.4934486} {\bibfield
  {journal} {\bibinfo  {journal} {Applied Physics Letters}\ }\textbf {\bibinfo
  {volume} {107}} (\bibinfo {year} {2015}),\ 10.1063/1.4934486},\ \Eprint
  {http://arxiv.org/abs/1509.01854} {arXiv:1509.01854} \BibitemShut {NoStop}%
\bibitem [{\citenamefont {M{\"{u}}ller}\ \emph {et~al.}(2019)\citenamefont
  {M{\"{u}}ller}, \citenamefont {Cole},\ and\ \citenamefont
  {Lisenfeld}}]{Muller2017}%
  \BibitemOpen
  \bibfield  {author} {\bibinfo {author} {\bibfnamefont {C.}~\bibnamefont
  {M{\"{u}}ller}}, \bibinfo {author} {\bibfnamefont {J.~H.}\ \bibnamefont
  {Cole}}, \ and\ \bibinfo {author} {\bibfnamefont {J.}~\bibnamefont
  {Lisenfeld}},\ }\href {\doibase 10.1088/1361-6633/ab3a7e} {\bibfield
  {journal} {\bibinfo  {journal} {Reports on Progress in Physics}\ }\textbf
  {\bibinfo {volume} {82}},\ \bibinfo {pages} {124501} (\bibinfo {year}
  {2019})},\ \Eprint {http://arxiv.org/abs/1705.01108} {arXiv:1705.01108}
  \BibitemShut {NoStop}%
\bibitem [{\citenamefont {Nersisyan}\ \emph {et~al.}(2019)\citenamefont
  {Nersisyan}, \citenamefont {Poletto}, \citenamefont {Alidoust}, \citenamefont
  {Manenti}, \citenamefont {Renzas}, \citenamefont {Bui}, \citenamefont {Vu},
  \citenamefont {Whyland}, \citenamefont {Mohan}, \citenamefont {Sete},
  \citenamefont {Stanwyck}, \citenamefont {Bestwick},\ and\ \citenamefont
  {Reagor}}]{Nersisyan2019}%
  \BibitemOpen
  \bibfield  {author} {\bibinfo {author} {\bibfnamefont {A.}~\bibnamefont
  {Nersisyan}}, \bibinfo {author} {\bibfnamefont {S.}~\bibnamefont {Poletto}},
  \bibinfo {author} {\bibfnamefont {N.}~\bibnamefont {Alidoust}}, \bibinfo
  {author} {\bibfnamefont {R.}~\bibnamefont {Manenti}}, \bibinfo {author}
  {\bibfnamefont {R.}~\bibnamefont {Renzas}}, \bibinfo {author} {\bibfnamefont
  {C.-V.}\ \bibnamefont {Bui}}, \bibinfo {author} {\bibfnamefont
  {K.}~\bibnamefont {Vu}}, \bibinfo {author} {\bibfnamefont {T.}~\bibnamefont
  {Whyland}}, \bibinfo {author} {\bibfnamefont {Y.}~\bibnamefont {Mohan}},
  \bibinfo {author} {\bibfnamefont {E.~A.}\ \bibnamefont {Sete}}, \bibinfo
  {author} {\bibfnamefont {S.}~\bibnamefont {Stanwyck}}, \bibinfo {author}
  {\bibfnamefont {A.}~\bibnamefont {Bestwick}}, \ and\ \bibinfo {author}
  {\bibfnamefont {M.}~\bibnamefont {Reagor}},\ }\href
  {http://arxiv.org/abs/1901.08042} {\  (\bibinfo {year} {2019})},\ \Eprint
  {http://arxiv.org/abs/1901.08042} {arXiv:1901.08042} \BibitemShut {NoStop}%
\bibitem [{\citenamefont {Dolan}(1977)}]{Dolan1977}%
  \BibitemOpen
  \bibfield  {author} {\bibinfo {author} {\bibfnamefont {G.~J.}\ \bibnamefont
  {Dolan}},\ }\href {\doibase 10.1063/1.89690} {\bibfield  {journal} {\bibinfo
  {journal} {Applied Physics Letters}\ }\textbf {\bibinfo {volume} {31}},\
  \bibinfo {pages} {337} (\bibinfo {year} {1977})}\BibitemShut {NoStop}%
\bibitem [{\citenamefont {Roddatis}\ \emph {et~al.}(2011)\citenamefont
  {Roddatis}, \citenamefont {H{\"{u}}bner}, \citenamefont {Ivanov},
  \citenamefont {Il'Ichev}, \citenamefont {Meyer}, \citenamefont {Koval'Chuk},\
  and\ \citenamefont {Vasiliev}}]{Roddatis2011}%
  \BibitemOpen
  \bibfield  {author} {\bibinfo {author} {\bibfnamefont {V.~V.}\ \bibnamefont
  {Roddatis}}, \bibinfo {author} {\bibfnamefont {U.}~\bibnamefont
  {H{\"{u}}bner}}, \bibinfo {author} {\bibfnamefont {B.~I.}\ \bibnamefont
  {Ivanov}}, \bibinfo {author} {\bibfnamefont {E.}~\bibnamefont {Il'Ichev}},
  \bibinfo {author} {\bibfnamefont {H.~G.}\ \bibnamefont {Meyer}}, \bibinfo
  {author} {\bibfnamefont {M.~V.}\ \bibnamefont {Koval'Chuk}}, \ and\ \bibinfo
  {author} {\bibfnamefont {A.~L.}\ \bibnamefont {Vasiliev}},\ }\href {\doibase
  10.1063/1.3670003} {\bibfield  {journal} {\bibinfo  {journal} {Journal of
  Applied Physics}\ }\textbf {\bibinfo {volume} {110}} (\bibinfo {year}
  {2011}),\ 10.1063/1.3670003}\BibitemShut {NoStop}%
\bibitem [{\citenamefont {Satoh}\ \emph {et~al.}(2015)\citenamefont {Satoh},
  \citenamefont {Noguchi}, \citenamefont {Yamagishi}, \citenamefont {Nagasawa},
  \citenamefont {Hinode}, \citenamefont {Hidaka}, \citenamefont {Maezawa},\
  and\ \citenamefont {Horikawa}}]{Satoh2015}%
  \BibitemOpen
  \bibfield  {author} {\bibinfo {author} {\bibfnamefont {T.}~\bibnamefont
  {Satoh}}, \bibinfo {author} {\bibfnamefont {Y.}~\bibnamefont {Noguchi}},
  \bibinfo {author} {\bibfnamefont {M.}~\bibnamefont {Yamagishi}}, \bibinfo
  {author} {\bibfnamefont {S.}~\bibnamefont {Nagasawa}}, \bibinfo {author}
  {\bibfnamefont {K.}~\bibnamefont {Hinode}}, \bibinfo {author} {\bibfnamefont
  {M.}~\bibnamefont {Hidaka}}, \bibinfo {author} {\bibfnamefont
  {M.}~\bibnamefont {Maezawa}}, \ and\ \bibinfo {author} {\bibfnamefont
  {T.}~\bibnamefont {Horikawa}},\ }\href {\doibase 10.1109/TASC.2014.2378033}
  {\bibfield  {journal} {\bibinfo  {journal} {IEEE Transactions on Applied
  Superconductivity}\ }\textbf {\bibinfo {volume} {25}} (\bibinfo {year}
  {2015}),\ 10.1109/TASC.2014.2378033}\BibitemShut {NoStop}%
\bibitem [{\citenamefont {Lecocq}\ \emph {et~al.}(2011)\citenamefont {Lecocq},
  \citenamefont {Pop}, \citenamefont {Peng}, \citenamefont {Matei},
  \citenamefont {Crozes}, \citenamefont {Fournier}, \citenamefont {Naud},
  \citenamefont {Guichard},\ and\ \citenamefont {Buisson}}]{Lecocq2011}%
  \BibitemOpen
  \bibfield  {author} {\bibinfo {author} {\bibfnamefont {F.}~\bibnamefont
  {Lecocq}}, \bibinfo {author} {\bibfnamefont {I.~M.}\ \bibnamefont {Pop}},
  \bibinfo {author} {\bibfnamefont {Z.}~\bibnamefont {Peng}}, \bibinfo {author}
  {\bibfnamefont {I.}~\bibnamefont {Matei}}, \bibinfo {author} {\bibfnamefont
  {T.}~\bibnamefont {Crozes}}, \bibinfo {author} {\bibfnamefont
  {T.}~\bibnamefont {Fournier}}, \bibinfo {author} {\bibfnamefont
  {C.}~\bibnamefont {Naud}}, \bibinfo {author} {\bibfnamefont {W.}~\bibnamefont
  {Guichard}}, \ and\ \bibinfo {author} {\bibfnamefont {O.}~\bibnamefont
  {Buisson}},\ }\href {\doibase 10.1088/0957-4484/22/31/315302} {\bibfield
  {journal} {\bibinfo  {journal} {Nanotechnology}\ }\textbf {\bibinfo {volume}
  {22}},\ \bibinfo {pages} {315302} (\bibinfo {year} {2011})}\BibitemShut
  {NoStop}%
\bibitem [{\citenamefont {Zhang}\ \emph {et~al.}(2017)\citenamefont {Zhang},
  \citenamefont {Li}, \citenamefont {Liu}, \citenamefont {Yu},\ and\
  \citenamefont {Yu}}]{Zhang2017}%
  \BibitemOpen
  \bibfield  {author} {\bibinfo {author} {\bibfnamefont {K.}~\bibnamefont
  {Zhang}}, \bibinfo {author} {\bibfnamefont {M.~M.}\ \bibnamefont {Li}},
  \bibinfo {author} {\bibfnamefont {Q.}~\bibnamefont {Liu}}, \bibinfo {author}
  {\bibfnamefont {H.~F.}\ \bibnamefont {Yu}}, \ and\ \bibinfo {author}
  {\bibfnamefont {Y.}~\bibnamefont {Yu}},\ }\href {\doibase
  10.1088/1674-1056/26/7/078501} {\bibfield  {journal} {\bibinfo  {journal}
  {Chinese Physics B}\ }\textbf {\bibinfo {volume} {26}},\ \bibinfo {pages} {0}
  (\bibinfo {year} {2017})}\BibitemShut {NoStop}%
\bibitem [{\citenamefont {Zeng}\ \emph
  {et~al.}(2015{\natexlab{a}})\citenamefont {Zeng}, \citenamefont {Nik},
  \citenamefont {Greibe}, \citenamefont {Krantz}, \citenamefont {Wilson},
  \citenamefont {Delsing},\ and\ \citenamefont {Olsson}}]{Zeng2015a}%
  \BibitemOpen
  \bibfield  {author} {\bibinfo {author} {\bibfnamefont {L.~J.}\ \bibnamefont
  {Zeng}}, \bibinfo {author} {\bibfnamefont {S.}~\bibnamefont {Nik}}, \bibinfo
  {author} {\bibfnamefont {T.}~\bibnamefont {Greibe}}, \bibinfo {author}
  {\bibfnamefont {P.}~\bibnamefont {Krantz}}, \bibinfo {author} {\bibfnamefont
  {C.~M.}\ \bibnamefont {Wilson}}, \bibinfo {author} {\bibfnamefont
  {P.}~\bibnamefont {Delsing}}, \ and\ \bibinfo {author} {\bibfnamefont
  {E.}~\bibnamefont {Olsson}},\ }\href {\doibase
  10.1088/0022-3727/48/39/395308} {\bibfield  {journal} {\bibinfo  {journal}
  {Journal of Physics D: Applied Physics}\ }\textbf {\bibinfo {volume} {48}},\
  \bibinfo {pages} {395308} (\bibinfo {year} {2015}{\natexlab{a}})}\BibitemShut
  {NoStop}%
\bibitem [{\citenamefont {Campbell}\ \emph {et~al.}(2005)\citenamefont
  {Campbell}, \citenamefont {Aral}, \citenamefont {Ogata}, \citenamefont
  {Kalia}, \citenamefont {Nakano},\ and\ \citenamefont
  {Vashishta}}]{Campbell2005}%
  \BibitemOpen
  \bibfield  {author} {\bibinfo {author} {\bibfnamefont {T.}~\bibnamefont
  {Campbell}}, \bibinfo {author} {\bibfnamefont {G.}~\bibnamefont {Aral}},
  \bibinfo {author} {\bibfnamefont {S.}~\bibnamefont {Ogata}}, \bibinfo
  {author} {\bibfnamefont {R.}~\bibnamefont {Kalia}}, \bibinfo {author}
  {\bibfnamefont {A.}~\bibnamefont {Nakano}}, \ and\ \bibinfo {author}
  {\bibfnamefont {P.}~\bibnamefont {Vashishta}},\ }\href {\doibase
  10.1103/PhysRevB.71.205413} {\bibfield  {journal} {\bibinfo  {journal}
  {Physical Review B}\ }\textbf {\bibinfo {volume} {71}},\ \bibinfo {pages}
  {205413} (\bibinfo {year} {2005})}\BibitemShut {NoStop}%
\bibitem [{\citenamefont {Zhou}\ and\ \citenamefont {Wadley}(2005)}]{Zhou2005}%
  \BibitemOpen
  \bibfield  {author} {\bibinfo {author} {\bibfnamefont {X.}~\bibnamefont
  {Zhou}}\ and\ \bibinfo {author} {\bibfnamefont {H.}~\bibnamefont {Wadley}},\
  }\href {\doibase 10.1103/PhysRevB.71.054418} {\bibfield  {journal} {\bibinfo
  {journal} {Physical Review B}\ }\textbf {\bibinfo {volume} {71}},\ \bibinfo
  {pages} {054418} (\bibinfo {year} {2005})}\BibitemShut {NoStop}%
\bibitem [{\citenamefont {Hasnaoui}\ \emph {et~al.}(2006)\citenamefont
  {Hasnaoui}, \citenamefont {Politano}, \citenamefont {Salazar},\ and\
  \citenamefont {Aral}}]{Hasnaoui2006}%
  \BibitemOpen
  \bibfield  {author} {\bibinfo {author} {\bibfnamefont {A.}~\bibnamefont
  {Hasnaoui}}, \bibinfo {author} {\bibfnamefont {O.}~\bibnamefont {Politano}},
  \bibinfo {author} {\bibfnamefont {J.~M.}\ \bibnamefont {Salazar}}, \ and\
  \bibinfo {author} {\bibfnamefont {G.}~\bibnamefont {Aral}},\ }\href {\doibase
  10.1103/PhysRevB.73.035427} {\bibfield  {journal} {\bibinfo  {journal}
  {Physical Review B}\ }\textbf {\bibinfo {volume} {73}},\ \bibinfo {pages}
  {035427} (\bibinfo {year} {2006})}\BibitemShut {NoStop}%
\bibitem [{\citenamefont {Zhou}\ \emph {et~al.}(2007)\citenamefont {Zhou},
  \citenamefont {Wadley},\ and\ \citenamefont {Wang}}]{Zhou2007}%
  \BibitemOpen
  \bibfield  {author} {\bibinfo {author} {\bibfnamefont {X.}~\bibnamefont
  {Zhou}}, \bibinfo {author} {\bibfnamefont {H.}~\bibnamefont {Wadley}}, \ and\
  \bibinfo {author} {\bibfnamefont {D.}~\bibnamefont {Wang}},\ }\href {\doibase
  10.1016/j.commatsci.2006.10.006} {\bibfield  {journal} {\bibinfo  {journal}
  {Computational Materials Science}\ }\textbf {\bibinfo {volume} {39}},\
  \bibinfo {pages} {794} (\bibinfo {year} {2007})}\BibitemShut {NoStop}%
\bibitem [{\citenamefont {Hong}\ and\ \citenamefont {van
  Duin}(2015)}]{Hong2015}%
  \BibitemOpen
  \bibfield  {author} {\bibinfo {author} {\bibfnamefont {S.}~\bibnamefont
  {Hong}}\ and\ \bibinfo {author} {\bibfnamefont {A.~C.}\ \bibnamefont {van
  Duin}},\ }\href {\doibase 10.1021/acs.jpcc.5b04650} {\bibfield  {journal}
  {\bibinfo  {journal} {The Journal of Physical Chemistry C}\ }\textbf
  {\bibinfo {volume} {119}},\ \bibinfo {pages} {17876} (\bibinfo {year}
  {2015})}\BibitemShut {NoStop}%
\bibitem [{\citenamefont {Tan}\ \emph {et~al.}(2005)\citenamefont {Tan},
  \citenamefont {Mather}, \citenamefont {Perrella}, \citenamefont {Read},\ and\
  \citenamefont {Buhrman}}]{Tan2005}%
  \BibitemOpen
  \bibfield  {author} {\bibinfo {author} {\bibfnamefont {E.}~\bibnamefont
  {Tan}}, \bibinfo {author} {\bibfnamefont {P.}~\bibnamefont {Mather}},
  \bibinfo {author} {\bibfnamefont {A.}~\bibnamefont {Perrella}}, \bibinfo
  {author} {\bibfnamefont {J.}~\bibnamefont {Read}}, \ and\ \bibinfo {author}
  {\bibfnamefont {R.}~\bibnamefont {Buhrman}},\ }\href {\doibase
  10.1103/PhysRevB.71.161401} {\bibfield  {journal} {\bibinfo  {journal}
  {Physical Review B}\ }\textbf {\bibinfo {volume} {71}},\ \bibinfo {pages}
  {161401} (\bibinfo {year} {2005})}\BibitemShut {NoStop}%
\bibitem [{\citenamefont {Holmqvist}\ \emph {et~al.}(2008)\citenamefont
  {Holmqvist}, \citenamefont {Meschke},\ and\ \citenamefont
  {Pekola}}]{Holmqvist2008}%
  \BibitemOpen
  \bibfield  {author} {\bibinfo {author} {\bibfnamefont {T.}~\bibnamefont
  {Holmqvist}}, \bibinfo {author} {\bibfnamefont {M.}~\bibnamefont {Meschke}},
  \ and\ \bibinfo {author} {\bibfnamefont {J.~P.}\ \bibnamefont {Pekola}},\
  }\href {\doibase 10.1116/1.2817629} {\bibfield  {journal} {\bibinfo
  {journal} {Journal of Vacuum Science \& Technology B}\ }\textbf {\bibinfo
  {volume} {26}},\ \bibinfo {pages} {28} (\bibinfo {year} {2008})}\BibitemShut
  {NoStop}%
\bibitem [{\citenamefont {Cai}\ \emph {et~al.}(2011)\citenamefont {Cai},
  \citenamefont {Zhou}, \citenamefont {M{\"{u}}ller},\ and\ \citenamefont
  {Starr}}]{Cai2011}%
  \BibitemOpen
  \bibfield  {author} {\bibinfo {author} {\bibfnamefont {N.}~\bibnamefont
  {Cai}}, \bibinfo {author} {\bibfnamefont {G.}~\bibnamefont {Zhou}}, \bibinfo
  {author} {\bibfnamefont {K.}~\bibnamefont {M{\"{u}}ller}}, \ and\ \bibinfo
  {author} {\bibfnamefont {D.~E.}\ \bibnamefont {Starr}},\ }\href {\doibase
  10.1103/PhysRevB.84.125445} {\bibfield  {journal} {\bibinfo  {journal}
  {Physical Review B}\ }\textbf {\bibinfo {volume} {84}},\ \bibinfo {pages} {1}
  (\bibinfo {year} {2011})}\BibitemShut {NoStop}%
\bibitem [{\citenamefont {Gutierrez}\ \emph {et~al.}(2000)\citenamefont
  {Gutierrez}, \citenamefont {Belonoshko}, \citenamefont {Ahuja},\ and\
  \citenamefont {Johansson}}]{Gutierrez2000}%
  \BibitemOpen
  \bibfield  {author} {\bibinfo {author} {\bibfnamefont {G.}~\bibnamefont
  {Gutierrez}}, \bibinfo {author} {\bibfnamefont {A.~B.}\ \bibnamefont
  {Belonoshko}}, \bibinfo {author} {\bibfnamefont {R.}~\bibnamefont {Ahuja}}, \
  and\ \bibinfo {author} {\bibfnamefont {B.}~\bibnamefont {Johansson}},\
  }\href@noop {} {\bibfield  {journal} {\bibinfo  {journal} {Physical Review
  E}\ }\textbf {\bibinfo {volume} {61}},\ \bibinfo {pages} {2723} (\bibinfo
  {year} {2000})}\BibitemShut {NoStop}%
\bibitem [{\citenamefont {Colleoni}\ \emph {et~al.}(2015)\citenamefont
  {Colleoni}, \citenamefont {Miceli},\ and\ \citenamefont
  {Pasquarello}}]{Colleoni2015}%
  \BibitemOpen
  \bibfield  {author} {\bibinfo {author} {\bibfnamefont {D.}~\bibnamefont
  {Colleoni}}, \bibinfo {author} {\bibfnamefont {G.}~\bibnamefont {Miceli}}, \
  and\ \bibinfo {author} {\bibfnamefont {A.}~\bibnamefont {Pasquarello}},\
  }\href {\doibase 10.1063/1.4936240} {\bibfield  {journal} {\bibinfo
  {journal} {Applied Physics Letters}\ }\textbf {\bibinfo {volume} {107}}
  (\bibinfo {year} {2015}),\ 10.1063/1.4936240}\BibitemShut {NoStop}%
\bibitem [{\citenamefont {DuBois}\ \emph {et~al.}(2016)\citenamefont {DuBois},
  \citenamefont {Cyster}, \citenamefont {Opletal}, \citenamefont {Russo},\ and\
  \citenamefont {Cole}}]{DuBois2016}%
  \BibitemOpen
  \bibfield  {author} {\bibinfo {author} {\bibfnamefont {T.~C.}\ \bibnamefont
  {DuBois}}, \bibinfo {author} {\bibfnamefont {M.~J.}\ \bibnamefont {Cyster}},
  \bibinfo {author} {\bibfnamefont {G.}~\bibnamefont {Opletal}}, \bibinfo
  {author} {\bibfnamefont {S.~P.}\ \bibnamefont {Russo}}, \ and\ \bibinfo
  {author} {\bibfnamefont {J.~H.}\ \bibnamefont {Cole}},\ }\href {\doibase
  10.1080/08927022.2015.1068941} {\bibfield  {journal} {\bibinfo  {journal}
  {Molecular Simulation}\ }\textbf {\bibinfo {volume} {42}},\ \bibinfo {pages}
  {542} (\bibinfo {year} {2016})}\BibitemShut {NoStop}%
\bibitem [{\citenamefont {Su}\ \emph {et~al.}(2017)\citenamefont {Su},
  \citenamefont {Liu}, \citenamefont {Xu}, \citenamefont {Deng}, \citenamefont
  {Li}, \citenamefont {Tian}, \citenamefont {Zhu}, \citenamefont {Zheng},
  \citenamefont {Lv},\ and\ \citenamefont {Zhao}}]{Su2017}%
  \BibitemOpen
  \bibfield  {author} {\bibinfo {author} {\bibfnamefont {F.~F.}\ \bibnamefont
  {Su}}, \bibinfo {author} {\bibfnamefont {W.~Y.}\ \bibnamefont {Liu}},
  \bibinfo {author} {\bibfnamefont {H.~K.}\ \bibnamefont {Xu}}, \bibinfo
  {author} {\bibfnamefont {H.}~\bibnamefont {Deng}}, \bibinfo {author}
  {\bibfnamefont {Z.~Y.}\ \bibnamefont {Li}}, \bibinfo {author} {\bibfnamefont
  {Y.}~\bibnamefont {Tian}}, \bibinfo {author} {\bibfnamefont {X.~B.}\
  \bibnamefont {Zhu}}, \bibinfo {author} {\bibfnamefont {D.~N.}\ \bibnamefont
  {Zheng}}, \bibinfo {author} {\bibfnamefont {L.}~\bibnamefont {Lv}}, \ and\
  \bibinfo {author} {\bibfnamefont {S.~P.}\ \bibnamefont {Zhao}},\ }\href
  {\doibase 10.1088/1674-1056/26/6/060308} {\bibfield  {journal} {\bibinfo
  {journal} {Chinese Physics B}\ }\textbf {\bibinfo {volume} {26}} (\bibinfo
  {year} {2017}),\ 10.1088/1674-1056/26/6/060308}\BibitemShut {NoStop}%
\bibitem [{\citenamefont {Park}\ \emph {et~al.}(2002)\citenamefont {Park},
  \citenamefont {Bae},\ and\ \citenamefont {Lee}}]{Park2002}%
  \BibitemOpen
  \bibfield  {author} {\bibinfo {author} {\bibfnamefont {B.~G.}\ \bibnamefont
  {Park}}, \bibinfo {author} {\bibfnamefont {J.~Y.}\ \bibnamefont {Bae}}, \
  and\ \bibinfo {author} {\bibfnamefont {T.~D.}\ \bibnamefont {Lee}},\ }\href
  {\doibase 10.1063/1.1447210} {\bibfield  {journal} {\bibinfo  {journal}
  {Journal of Applied Physics}\ }\textbf {\bibinfo {volume} {91}},\ \bibinfo
  {pages} {8789} (\bibinfo {year} {2002})}\BibitemShut {NoStop}%
\bibitem [{\citenamefont {Suzuki}\ \emph {et~al.}(1993)\citenamefont {Suzuki},
  \citenamefont {Matsui}, \citenamefont {Ohi}, \citenamefont {Ishi},
  \citenamefont {Kimura}, \citenamefont {Tsuda},\ and\ \citenamefont
  {Mukae}}]{Suzuki1993}%
  \BibitemOpen
  \bibfield  {author} {\bibinfo {author} {\bibfnamefont {T.}~\bibnamefont
  {Suzuki}}, \bibinfo {author} {\bibfnamefont {T.}~\bibnamefont {Matsui}},
  \bibinfo {author} {\bibfnamefont {A.}~\bibnamefont {Ohi}}, \bibinfo {author}
  {\bibfnamefont {T.}~\bibnamefont {Ishi}}, \bibinfo {author} {\bibfnamefont
  {H.}~\bibnamefont {Kimura}}, \bibinfo {author} {\bibfnamefont
  {K.}~\bibnamefont {Tsuda}}, \ and\ \bibinfo {author} {\bibfnamefont
  {K.}~\bibnamefont {Mukae}},\ }in\ \href {\doibase
  10.1007/978-4-431-68305-6_215} {\emph {\bibinfo {booktitle} {Advances in
  Superconductivity V}}}\ (\bibinfo  {publisher} {Springer Japan},\ \bibinfo
  {address} {Tokyo},\ \bibinfo {year} {1993})\ pp.\ \bibinfo {pages}
  {961--964}\BibitemShut {NoStop}%
\bibitem [{\citenamefont {Roos}\ \emph {et~al.}(2001)\citenamefont {Roos},
  \citenamefont {Beck}, \citenamefont {Demokritov}, \citenamefont
  {Hillebrands},\ and\ \citenamefont {Ozkaya}}]{Roos2001}%
  \BibitemOpen
  \bibfield  {author} {\bibinfo {author} {\bibfnamefont {B.~F.}\ \bibnamefont
  {Roos}}, \bibinfo {author} {\bibfnamefont {P.~A.}\ \bibnamefont {Beck}},
  \bibinfo {author} {\bibfnamefont {S.~O.}\ \bibnamefont {Demokritov}},
  \bibinfo {author} {\bibfnamefont {B.}~\bibnamefont {Hillebrands}}, \ and\
  \bibinfo {author} {\bibfnamefont {D.}~\bibnamefont {Ozkaya}},\ }\href
  {\doibase 10.1063/1.1356709} {\bibfield  {journal} {\bibinfo  {journal}
  {Journal of Applied Physics}\ }\textbf {\bibinfo {volume} {89}},\ \bibinfo
  {pages} {6656} (\bibinfo {year} {2001})}\BibitemShut {NoStop}%
\bibitem [{\citenamefont {Schuhl}\ \emph {et~al.}(1990)\citenamefont {Schuhl},
  \citenamefont {Cabanel}, \citenamefont {Lequien}, \citenamefont {Ghyselen},
  \citenamefont {Ty{\v{c}}}, \citenamefont {Creuzet},\ and\ \citenamefont
  {Siejka}}]{Schuhl1990}%
  \BibitemOpen
  \bibfield  {author} {\bibinfo {author} {\bibfnamefont {A.}~\bibnamefont
  {Schuhl}}, \bibinfo {author} {\bibfnamefont {R.}~\bibnamefont {Cabanel}},
  \bibinfo {author} {\bibfnamefont {S.}~\bibnamefont {Lequien}}, \bibinfo
  {author} {\bibfnamefont {B.}~\bibnamefont {Ghyselen}}, \bibinfo {author}
  {\bibfnamefont {S.}~\bibnamefont {Ty{\v{c}}}}, \bibinfo {author}
  {\bibfnamefont {G.}~\bibnamefont {Creuzet}}, \ and\ \bibinfo {author}
  {\bibfnamefont {J.}~\bibnamefont {Siejka}},\ }\href {\doibase
  10.1063/1.103430} {\bibfield  {journal} {\bibinfo  {journal} {Applied Physics
  Letters}\ }\textbf {\bibinfo {volume} {57}},\ \bibinfo {pages} {819}
  (\bibinfo {year} {1990})}\BibitemShut {NoStop}%
\bibitem [{\citenamefont {Kleiman}\ \emph {et~al.}(2003)\citenamefont
  {Kleiman}, \citenamefont {Iskanderova}, \citenamefont {Gudimenko},\ and\
  \citenamefont {Horodetsky}}]{Kleiman2003}%
  \BibitemOpen
  \bibfield  {author} {\bibinfo {author} {\bibfnamefont {J.}~\bibnamefont
  {Kleiman}}, \bibinfo {author} {\bibfnamefont {Z.}~\bibnamefont
  {Iskanderova}}, \bibinfo {author} {\bibfnamefont {Y.}~\bibnamefont
  {Gudimenko}}, \ and\ \bibinfo {author} {\bibfnamefont {S.}~\bibnamefont
  {Horodetsky}},\ }in\ \href@noop {} {\emph {\bibinfo {booktitle} {European
  Space Agency, (Special Publication) ESA SP}}}\ (\bibinfo {year}
  {2003})\BibitemShut {NoStop}%
\bibitem [{\citenamefont {Brune}\ \emph {et~al.}(1992)\citenamefont {Brune},
  \citenamefont {Wintterlin}, \citenamefont {Behm},\ and\ \citenamefont
  {Ertl}}]{Brune1992}%
  \BibitemOpen
  \bibfield  {author} {\bibinfo {author} {\bibfnamefont {H.}~\bibnamefont
  {Brune}}, \bibinfo {author} {\bibfnamefont {J.}~\bibnamefont {Wintterlin}},
  \bibinfo {author} {\bibfnamefont {R.~J.}\ \bibnamefont {Behm}}, \ and\
  \bibinfo {author} {\bibfnamefont {G.}~\bibnamefont {Ertl}},\ }\href {\doibase
  10.1103/PhysRevLett.68.624} {\bibfield  {journal} {\bibinfo  {journal}
  {Physical Review Letters}\ }\textbf {\bibinfo {volume} {68}},\ \bibinfo
  {pages} {624} (\bibinfo {year} {1992})}\BibitemShut {NoStop}%
\bibitem [{\citenamefont {Gale}\ and\ \citenamefont {Rohl}(2003)}]{Gale2003}%
  \BibitemOpen
  \bibfield  {author} {\bibinfo {author} {\bibfnamefont {J.~D.}\ \bibnamefont
  {Gale}}\ and\ \bibinfo {author} {\bibfnamefont {A.~L.}\ \bibnamefont
  {Rohl}},\ }\href {\doibase 10.1080/0892702031000104887} {\bibfield  {journal}
  {\bibinfo  {journal} {Molecular Simulation}\ }\textbf {\bibinfo {volume}
  {29}},\ \bibinfo {pages} {291} (\bibinfo {year} {2003})},\ \Eprint
  {http://arxiv.org/abs/1011.1669v3} {arXiv:1011.1669v3} \BibitemShut {NoStop}%
\bibitem [{\citenamefont {Streitz}\ and\ \citenamefont
  {Mintmire}(1994)}]{Streitz1994}%
  \BibitemOpen
  \bibfield  {author} {\bibinfo {author} {\bibfnamefont {F.~H.}\ \bibnamefont
  {Streitz}}\ and\ \bibinfo {author} {\bibfnamefont {J.~W.}\ \bibnamefont
  {Mintmire}},\ }\href {\doibase 10.1103/PhysRevB.50.11996} {\bibfield
  {journal} {\bibinfo  {journal} {Physical Review B}\ }\textbf {\bibinfo
  {volume} {50}},\ \bibinfo {pages} {11996} (\bibinfo {year}
  {1994})}\BibitemShut {NoStop}%
\bibitem [{\citenamefont {Nos{\'{e}}}(1984)}]{Nose1984}%
  \BibitemOpen
  \bibfield  {author} {\bibinfo {author} {\bibfnamefont {S.}~\bibnamefont
  {Nos{\'{e}}}},\ }\href {\doibase 10.1080/00268978400101201} {\bibfield
  {journal} {\bibinfo  {journal} {Molecular Physics}\ }\textbf {\bibinfo
  {volume} {52}},\ \bibinfo {pages} {255} (\bibinfo {year} {1984})}\BibitemShut
  {NoStop}%
\bibitem [{\citenamefont {Hoover}(1985)}]{Hoover1985}%
  \BibitemOpen
  \bibfield  {author} {\bibinfo {author} {\bibfnamefont {W.~G.}\ \bibnamefont
  {Hoover}},\ }\href {\doibase 10.1103/PhysRevA.31.1695} {\bibfield  {journal}
  {\bibinfo  {journal} {Physical Review A}\ }\textbf {\bibinfo {volume} {31}},\
  \bibinfo {pages} {1695} (\bibinfo {year} {1985})}\BibitemShut {NoStop}%
\bibitem [{\citenamefont {Plimpton}(1995{\natexlab{a}})}]{Plimpton1995}%
  \BibitemOpen
  \bibfield  {author} {\bibinfo {author} {\bibfnamefont {S.}~\bibnamefont
  {Plimpton}},\ }\href {\doibase 10.1006/jcph.1995.1039} {\bibfield  {journal}
  {\bibinfo  {journal} {Journal of Computational Physics}\ }\textbf {\bibinfo
  {volume} {117}},\ \bibinfo {pages} {1} (\bibinfo {year}
  {1995}{\natexlab{a}})}\BibitemShut {NoStop}%
\bibitem [{\citenamefont {Plimpton}(1995{\natexlab{b}})}]{LAMMPS}%
  \BibitemOpen
  \bibfield  {author} {\bibinfo {author} {\bibfnamefont {S.}~\bibnamefont
  {Plimpton}},\ }\href {http://lammps.sandia.gov} {\enquote {\bibinfo {title}
  {{{LAMMPS}}},}\ }\bibinfo {howpublished} {\url{http://lammps.sandia.gov}}
  (\bibinfo {year} {1995}{\natexlab{b}})\BibitemShut {NoStop}%
\bibitem [{\citenamefont {{Van Duin}}\ \emph {et~al.}(2001)\citenamefont {{Van
  Duin}}, \citenamefont {Dasgupta}, \citenamefont {Lorant},\ and\ \citenamefont
  {Goddard}}]{vanDuin2001}%
  \BibitemOpen
  \bibfield  {author} {\bibinfo {author} {\bibfnamefont {A.~C.}\ \bibnamefont
  {{Van Duin}}}, \bibinfo {author} {\bibfnamefont {S.}~\bibnamefont
  {Dasgupta}}, \bibinfo {author} {\bibfnamefont {F.}~\bibnamefont {Lorant}}, \
  and\ \bibinfo {author} {\bibfnamefont {W.~A.}\ \bibnamefont {Goddard}},\
  }\href {\doibase 10.1021/jp004368u} {\bibfield  {journal} {\bibinfo
  {journal} {Journal of Physical Chemistry A}\ }\textbf {\bibinfo {volume}
  {105}},\ \bibinfo {pages} {9396} (\bibinfo {year} {2001})},\ \Eprint
  {http://arxiv.org/abs/1011.1669v3} {arXiv:1011.1669v3} \BibitemShut {NoStop}%
\bibitem [{\citenamefont {Aktulga}\ \emph {et~al.}(2012)\citenamefont
  {Aktulga}, \citenamefont {Fogarty}, \citenamefont {Pandit},\ and\
  \citenamefont {Grama}}]{Aktulga2012}%
  \BibitemOpen
  \bibfield  {author} {\bibinfo {author} {\bibfnamefont {H.~M.}\ \bibnamefont
  {Aktulga}}, \bibinfo {author} {\bibfnamefont {J.~C.}\ \bibnamefont
  {Fogarty}}, \bibinfo {author} {\bibfnamefont {S.~A.}\ \bibnamefont {Pandit}},
  \ and\ \bibinfo {author} {\bibfnamefont {A.~Y.}\ \bibnamefont {Grama}},\
  }\href {\doibase 10.1016/j.parco.2011.08.005} {\bibfield  {journal} {\bibinfo
   {journal} {Parallel Computing}\ }\textbf {\bibinfo {volume} {38}},\ \bibinfo
  {pages} {245} (\bibinfo {year} {2012})}\BibitemShut {NoStop}%
\bibitem [{\citenamefont {Cooper}(1962)}]{Cooper1962}%
  \BibitemOpen
  \bibfield  {author} {\bibinfo {author} {\bibfnamefont {A.~S.}\ \bibnamefont
  {Cooper}},\ }\href {\doibase 10.1107/S0365110X62001474} {\bibfield  {journal}
  {\bibinfo  {journal} {Acta Crystallographica}\ }\textbf {\bibinfo {volume}
  {15}},\ \bibinfo {pages} {578} (\bibinfo {year} {1962})}\BibitemShut
  {NoStop}%
\bibitem [{\citenamefont {Lide}(2005)}]{CRC}%
  \BibitemOpen
  \bibinfo {editor} {\bibfnamefont {D.~R.}\ \bibnamefont {Lide}},\ ed.,\ \href
  {https://s3.amazonaws.com/academia.edu.documents/49221189/CRC_Handbook_of_Chemistry_and_Physics_85th_ed_-_David_R._Lide.pdf?AWSAccessKeyId=AKIAIWOWYYGZ2Y53UL3A&Expires=1555380295&Signature=7gRYikJwdLM7m69sY%2FjpnJaD%2Bug%3D&response-content-disposition=inl}
  {\emph {\bibinfo {title} {{CRC Handbook of Chemistry and Physics}}}},\
  \bibinfo {edition} {85th}\ ed.\ (\bibinfo  {publisher} {CRC Press},\ \bibinfo
  {address} {Boca Raton, FL},\ \bibinfo {year} {2005})\BibitemShut {NoStop}%
\bibitem [{\citenamefont {Zeng}\ \emph {et~al.}(2016)\citenamefont {Zeng},
  \citenamefont {Tran}, \citenamefont {Tai}, \citenamefont {Svensson},\ and\
  \citenamefont {Olsson}}]{Zeng2016}%
  \BibitemOpen
  \bibfield  {author} {\bibinfo {author} {\bibfnamefont {L.}~\bibnamefont
  {Zeng}}, \bibinfo {author} {\bibfnamefont {D.~T.}\ \bibnamefont {Tran}},
  \bibinfo {author} {\bibfnamefont {C.-w.}\ \bibnamefont {Tai}}, \bibinfo
  {author} {\bibfnamefont {G.}~\bibnamefont {Svensson}}, \ and\ \bibinfo
  {author} {\bibfnamefont {E.}~\bibnamefont {Olsson}},\ }\href {\doibase
  10.1038/srep29679} {\bibfield  {journal} {\bibinfo  {journal} {Scientific
  Reports}\ }\textbf {\bibinfo {volume} {6}},\ \bibinfo {pages} {29679}
  (\bibinfo {year} {2016})}\BibitemShut {NoStop}%
\bibitem [{\citenamefont {Opletal}\ \emph {et~al.}(2019)\citenamefont
  {Opletal}, \citenamefont {Sun}, \citenamefont {Petersen}, \citenamefont
  {Russo},\ and\ \citenamefont {Barnard}}]{Opletal2019}%
  \BibitemOpen
  \bibfield  {author} {\bibinfo {author} {\bibfnamefont {G.}~\bibnamefont
  {Opletal}}, \bibinfo {author} {\bibfnamefont {B.}~\bibnamefont {Sun}},
  \bibinfo {author} {\bibfnamefont {T.~C.}\ \bibnamefont {Petersen}}, \bibinfo
  {author} {\bibfnamefont {S.~P.}\ \bibnamefont {Russo}}, \ and\ \bibinfo
  {author} {\bibfnamefont {A.~S.}\ \bibnamefont {Barnard}},\ }\href {\doibase
  10.1039/c8cp06649c} {\bibfield  {journal} {\bibinfo  {journal} {Physical
  Chemistry Chemical Physics}\ }\textbf {\bibinfo {volume} {21}},\ \bibinfo
  {pages} {6517} (\bibinfo {year} {2019})}\BibitemShut {NoStop}%
\bibitem [{\citenamefont {Cabrera}\ and\ \citenamefont
  {Mott}(1948)}]{Cabrera1948}%
  \BibitemOpen
  \bibfield  {author} {\bibinfo {author} {\bibfnamefont {N.}~\bibnamefont
  {Cabrera}}\ and\ \bibinfo {author} {\bibfnamefont {N.~F.}\ \bibnamefont
  {Mott}},\ }\href {\doibase 10.1088/0034-4885/12/1/308} {\bibfield  {journal}
  {\bibinfo  {journal} {Reports on Progress in Physics}\ }\textbf {\bibinfo
  {volume} {12}},\ \bibinfo {pages} {163} (\bibinfo {year} {1948})},\ \Eprint
  {http://arxiv.org/abs/1011.1669v3} {arXiv:1011.1669v3} \BibitemShut {NoStop}%
\bibitem [{\citenamefont {Nguyen}\ \emph {et~al.}(2018)\citenamefont {Nguyen},
  \citenamefont {Hashimoto}, \citenamefont {Zakharov}, \citenamefont {Stach},
  \citenamefont {Rooney}, \citenamefont {Berkels}, \citenamefont {Thompson},
  \citenamefont {Haigh},\ and\ \citenamefont {Burnett}}]{Nguyen2018}%
  \BibitemOpen
  \bibfield  {author} {\bibinfo {author} {\bibfnamefont {L.}~\bibnamefont
  {Nguyen}}, \bibinfo {author} {\bibfnamefont {T.}~\bibnamefont {Hashimoto}},
  \bibinfo {author} {\bibfnamefont {D.~N.}\ \bibnamefont {Zakharov}}, \bibinfo
  {author} {\bibfnamefont {E.~A.}\ \bibnamefont {Stach}}, \bibinfo {author}
  {\bibfnamefont {A.~P.}\ \bibnamefont {Rooney}}, \bibinfo {author}
  {\bibfnamefont {B.}~\bibnamefont {Berkels}}, \bibinfo {author} {\bibfnamefont
  {G.~E.}\ \bibnamefont {Thompson}}, \bibinfo {author} {\bibfnamefont {S.~J.}\
  \bibnamefont {Haigh}}, \ and\ \bibinfo {author} {\bibfnamefont {T.~L.}\
  \bibnamefont {Burnett}},\ }\href {\doibase 10.1021/acsami.7b17224} {\bibfield
   {journal} {\bibinfo  {journal} {ACS Applied Materials and Interfaces}\
  }\textbf {\bibinfo {volume} {10}},\ \bibinfo {pages} {2230} (\bibinfo {year}
  {2018})}\BibitemShut {NoStop}%
\bibitem [{\citenamefont {Jeurgens}\ \emph
  {et~al.}(2002{\natexlab{a}})\citenamefont {Jeurgens}, \citenamefont {Sloof},
  \citenamefont {Tichelaar},\ and\ \citenamefont
  {Mittemeijer}}]{Jeurgens2002b}%
  \BibitemOpen
  \bibfield  {author} {\bibinfo {author} {\bibfnamefont {L.~P.}\ \bibnamefont
  {Jeurgens}}, \bibinfo {author} {\bibfnamefont {W.~G.}\ \bibnamefont {Sloof}},
  \bibinfo {author} {\bibfnamefont {F.~D.}\ \bibnamefont {Tichelaar}}, \ and\
  \bibinfo {author} {\bibfnamefont {E.~J.}\ \bibnamefont {Mittemeijer}},\
  }\href {\doibase 10.1016/S0039-6028(02)01432-2} {\bibfield  {journal}
  {\bibinfo  {journal} {Surface Science}\ }\textbf {\bibinfo {volume} {506}},\
  \bibinfo {pages} {313} (\bibinfo {year} {2002}{\natexlab{a}})}\BibitemShut
  {NoStop}%
\bibitem [{\citenamefont {Aref}\ \emph {et~al.}(2014)\citenamefont {Aref},
  \citenamefont {Averin}, \citenamefont {van Dijken}, \citenamefont {Ferring},
  \citenamefont {Koberidze}, \citenamefont {Maisi}, \citenamefont {Nguyend},
  \citenamefont {Nieminen}, \citenamefont {Pekola},\ and\ \citenamefont
  {Yao}}]{Aref2014}%
  \BibitemOpen
  \bibfield  {author} {\bibinfo {author} {\bibfnamefont {T.}~\bibnamefont
  {Aref}}, \bibinfo {author} {\bibfnamefont {A.}~\bibnamefont {Averin}},
  \bibinfo {author} {\bibfnamefont {S.}~\bibnamefont {van Dijken}}, \bibinfo
  {author} {\bibfnamefont {A.}~\bibnamefont {Ferring}}, \bibinfo {author}
  {\bibfnamefont {M.}~\bibnamefont {Koberidze}}, \bibinfo {author}
  {\bibfnamefont {V.~F.}\ \bibnamefont {Maisi}}, \bibinfo {author}
  {\bibfnamefont {H.~Q.}\ \bibnamefont {Nguyend}}, \bibinfo {author}
  {\bibfnamefont {R.~M.}\ \bibnamefont {Nieminen}}, \bibinfo {author}
  {\bibfnamefont {J.~P.}\ \bibnamefont {Pekola}}, \ and\ \bibinfo {author}
  {\bibfnamefont {L.~D.}\ \bibnamefont {Yao}},\ }\href {\doibase
  10.1063/1.4893473} {\bibfield  {journal} {\bibinfo  {journal} {Journal of
  Applied Physics}\ }\textbf {\bibinfo {volume} {116}},\ \bibinfo {pages}
  {073702} (\bibinfo {year} {2014})}\BibitemShut {NoStop}%
\bibitem [{\citenamefont {Nik}\ \emph {et~al.}(2016)\citenamefont {Nik},
  \citenamefont {Krantz}, \citenamefont {Zeng}, \citenamefont {Greibe},
  \citenamefont {Pettersson}, \citenamefont {Gustafsson}, \citenamefont
  {Delsing},\ and\ \citenamefont {Olsson}}]{Nik2016}%
  \BibitemOpen
  \bibfield  {author} {\bibinfo {author} {\bibfnamefont {S.}~\bibnamefont
  {Nik}}, \bibinfo {author} {\bibfnamefont {P.}~\bibnamefont {Krantz}},
  \bibinfo {author} {\bibfnamefont {L.}~\bibnamefont {Zeng}}, \bibinfo {author}
  {\bibfnamefont {T.}~\bibnamefont {Greibe}}, \bibinfo {author} {\bibfnamefont
  {H.}~\bibnamefont {Pettersson}}, \bibinfo {author} {\bibfnamefont
  {S.}~\bibnamefont {Gustafsson}}, \bibinfo {author} {\bibfnamefont
  {P.}~\bibnamefont {Delsing}}, \ and\ \bibinfo {author} {\bibfnamefont
  {E.}~\bibnamefont {Olsson}},\ }\href {\doibase 10.1186/s40064-016-2418-8}
  {\bibfield  {journal} {\bibinfo  {journal} {SpringerPlus}\ }\textbf {\bibinfo
  {volume} {5}} (\bibinfo {year} {2016}),\
  10.1186/s40064-016-2418-8}\BibitemShut {NoStop}%
\bibitem [{\citenamefont {Fritz}\ \emph {et~al.}(2019)\citenamefont {Fritz},
  \citenamefont {Radtke}, \citenamefont {Schneider}, \citenamefont {Weides},\
  and\ \citenamefont {Gerthsen}}]{Fritz2019}%
  \BibitemOpen
  \bibfield  {author} {\bibinfo {author} {\bibfnamefont {S.}~\bibnamefont
  {Fritz}}, \bibinfo {author} {\bibfnamefont {L.}~\bibnamefont {Radtke}},
  \bibinfo {author} {\bibfnamefont {R.}~\bibnamefont {Schneider}}, \bibinfo
  {author} {\bibfnamefont {M.}~\bibnamefont {Weides}}, \ and\ \bibinfo {author}
  {\bibfnamefont {D.}~\bibnamefont {Gerthsen}},\ }\href {\doibase
  10.1063/1.5089871} {\bibfield  {journal} {\bibinfo  {journal} {Journal of
  Applied Physics}\ }\textbf {\bibinfo {volume} {125}} (\bibinfo {year}
  {2019}),\ 10.1063/1.5089871}\BibitemShut {NoStop}%
\bibitem [{\citenamefont {Jeurgens}\ \emph {et~al.}(1999)\citenamefont
  {Jeurgens}, \citenamefont {Sloof}, \citenamefont {Tichelaar}, \citenamefont
  {Borsboom},\ and\ \citenamefont {Mittemeijer}}]{Jeurgens1999}%
  \BibitemOpen
  \bibfield  {author} {\bibinfo {author} {\bibfnamefont {L.~P.}\ \bibnamefont
  {Jeurgens}}, \bibinfo {author} {\bibfnamefont {W.~G.}\ \bibnamefont {Sloof}},
  \bibinfo {author} {\bibfnamefont {F.~D.}\ \bibnamefont {Tichelaar}}, \bibinfo
  {author} {\bibfnamefont {C.~G.}\ \bibnamefont {Borsboom}}, \ and\ \bibinfo
  {author} {\bibfnamefont {E.~J.}\ \bibnamefont {Mittemeijer}},\ }\href
  {\doibase 10.1016/S0169-4332(98)00755-7} {\bibfield  {journal} {\bibinfo
  {journal} {Applied Surface Science}\ }\textbf {\bibinfo {volume} {144-145}},\
  \bibinfo {pages} {11} (\bibinfo {year} {1999})}\BibitemShut {NoStop}%
\bibitem [{\citenamefont {Reichel}\ \emph
  {et~al.}(2008{\natexlab{a}})\citenamefont {Reichel}, \citenamefont
  {Jeurgens}, \citenamefont {Richter},\ and\ \citenamefont
  {Mittemeijer}}]{Reichel2008}%
  \BibitemOpen
  \bibfield  {author} {\bibinfo {author} {\bibfnamefont {F.}~\bibnamefont
  {Reichel}}, \bibinfo {author} {\bibfnamefont {L.~P.~H.}\ \bibnamefont
  {Jeurgens}}, \bibinfo {author} {\bibfnamefont {G.}~\bibnamefont {Richter}}, \
  and\ \bibinfo {author} {\bibfnamefont {E.~J.}\ \bibnamefont {Mittemeijer}},\
  }\href {\doibase 10.1063/1.2913505} {\bibfield  {journal} {\bibinfo
  {journal} {Journal of Applied Physics}\ }\textbf {\bibinfo {volume} {103}}
  (\bibinfo {year} {2008}{\natexlab{a}}),\ 10.1063/1.2913505}\BibitemShut
  {NoStop}%
\bibitem [{\citenamefont {Jeurgens}\ \emph
  {et~al.}(2002{\natexlab{b}})\citenamefont {Jeurgens}, \citenamefont {Sloof},
  \citenamefont {Tichelaar},\ and\ \citenamefont
  {Mittemeijer}}]{Jeurgens2002c}%
  \BibitemOpen
  \bibfield  {author} {\bibinfo {author} {\bibfnamefont {L.~P.}\ \bibnamefont
  {Jeurgens}}, \bibinfo {author} {\bibfnamefont {W.~G.}\ \bibnamefont {Sloof}},
  \bibinfo {author} {\bibfnamefont {F.~D.}\ \bibnamefont {Tichelaar}}, \ and\
  \bibinfo {author} {\bibfnamefont {E.~J.}\ \bibnamefont {Mittemeijer}},\
  }\href {\doibase 10.1063/1.1491591} {\bibfield  {journal} {\bibinfo
  {journal} {Journal of Applied Physics}\ }\textbf {\bibinfo {volume} {92}},\
  \bibinfo {pages} {1649} (\bibinfo {year} {2002}{\natexlab{b}})}\BibitemShut
  {NoStop}%
\bibitem [{\citenamefont {Fl{\"{o}}totto}\ \emph {et~al.}(2015)\citenamefont
  {Fl{\"{o}}totto}, \citenamefont {Wang},\ and\ \citenamefont
  {Mittemeijer}}]{Flototto2015}%
  \BibitemOpen
  \bibfield  {author} {\bibinfo {author} {\bibfnamefont {D.}~\bibnamefont
  {Fl{\"{o}}totto}}, \bibinfo {author} {\bibfnamefont {Z.~M.}\ \bibnamefont
  {Wang}}, \ and\ \bibinfo {author} {\bibfnamefont {E.~J.}\ \bibnamefont
  {Mittemeijer}},\ }\href {\doibase 10.1016/j.susc.2014.11.008} {\bibfield
  {journal} {\bibinfo  {journal} {Surface Science}\ }\textbf {\bibinfo {volume}
  {633}},\ \bibinfo {pages} {1} (\bibinfo {year} {2015})}\BibitemShut {NoStop}%
\bibitem [{\citenamefont {Sankaranarayanan}\ \emph {et~al.}(2009)\citenamefont
  {Sankaranarayanan}, \citenamefont {Kaxiras},\ and\ \citenamefont
  {Ramanathan}}]{Sankaranarayanan2009}%
  \BibitemOpen
  \bibfield  {author} {\bibinfo {author} {\bibfnamefont {S.}~\bibnamefont
  {Sankaranarayanan}}, \bibinfo {author} {\bibfnamefont {E.}~\bibnamefont
  {Kaxiras}}, \ and\ \bibinfo {author} {\bibfnamefont {S.}~\bibnamefont
  {Ramanathan}},\ }\href {\doibase 10.1103/PhysRevLett.102.095504} {\bibfield
  {journal} {\bibinfo  {journal} {Physical Review Letters}\ }\textbf {\bibinfo
  {volume} {102}},\ \bibinfo {pages} {095504} (\bibinfo {year}
  {2009})}\BibitemShut {NoStop}%
\bibitem [{\citenamefont {Fritz}\ \emph {et~al.}(2018)\citenamefont {Fritz},
  \citenamefont {Seiler}, \citenamefont {Radtke}, \citenamefont {Schneider},
  \citenamefont {Weides}, \citenamefont {Wei{\ss}},\ and\ \citenamefont
  {Gerthsen}}]{Fritz2018}%
  \BibitemOpen
  \bibfield  {author} {\bibinfo {author} {\bibfnamefont {S.}~\bibnamefont
  {Fritz}}, \bibinfo {author} {\bibfnamefont {A.}~\bibnamefont {Seiler}},
  \bibinfo {author} {\bibfnamefont {L.}~\bibnamefont {Radtke}}, \bibinfo
  {author} {\bibfnamefont {R.}~\bibnamefont {Schneider}}, \bibinfo {author}
  {\bibfnamefont {M.}~\bibnamefont {Weides}}, \bibinfo {author} {\bibfnamefont
  {G.}~\bibnamefont {Wei{\ss}}}, \ and\ \bibinfo {author} {\bibfnamefont
  {D.}~\bibnamefont {Gerthsen}},\ }\href {\doibase 10.1038/s41598-018-26066-4}
  {\bibfield  {journal} {\bibinfo  {journal} {Scientific Reports}\ }\textbf
  {\bibinfo {volume} {8}},\ \bibinfo {pages} {1} (\bibinfo {year} {2018})},\
  \Eprint {http://arxiv.org/abs/1712.01581} {arXiv:1712.01581} \BibitemShut
  {NoStop}%
\bibitem [{\citenamefont {Jeurgens}\ \emph
  {et~al.}(2002{\natexlab{c}})\citenamefont {Jeurgens}, \citenamefont {Sloof},
  \citenamefont {Tichelaar},\ and\ \citenamefont
  {Mittemeijer}}]{Jeurgens2002a}%
  \BibitemOpen
  \bibfield  {author} {\bibinfo {author} {\bibfnamefont {L.}~\bibnamefont
  {Jeurgens}}, \bibinfo {author} {\bibfnamefont {W.}~\bibnamefont {Sloof}},
  \bibinfo {author} {\bibfnamefont {F.}~\bibnamefont {Tichelaar}}, \ and\
  \bibinfo {author} {\bibfnamefont {E.}~\bibnamefont {Mittemeijer}},\ }\href
  {\doibase 10.1016/S0040-6090(02)00787-3} {\bibfield  {journal} {\bibinfo
  {journal} {Thin Solid Films}\ }\textbf {\bibinfo {volume} {418}},\ \bibinfo
  {pages} {89} (\bibinfo {year} {2002}{\natexlab{c}})}\BibitemShut {NoStop}%
\bibitem [{\citenamefont {Reichel}\ \emph
  {et~al.}(2008{\natexlab{b}})\citenamefont {Reichel}, \citenamefont
  {Jeurgens}, \citenamefont {Richter},\ and\ \citenamefont
  {Mittemeijer}}]{Reichel2008a}%
  \BibitemOpen
  \bibfield  {author} {\bibinfo {author} {\bibfnamefont {F.}~\bibnamefont
  {Reichel}}, \bibinfo {author} {\bibfnamefont {L.~P.~H.}\ \bibnamefont
  {Jeurgens}}, \bibinfo {author} {\bibfnamefont {G.}~\bibnamefont {Richter}}, \
  and\ \bibinfo {author} {\bibfnamefont {E.~J.}\ \bibnamefont {Mittemeijer}},\
  }\href {\doibase 10.1063/1.2913505} {\bibfield  {journal} {\bibinfo
  {journal} {Journal of Applied Physics}\ }\textbf {\bibinfo {volume} {103}}
  (\bibinfo {year} {2008}{\natexlab{b}}),\ 10.1063/1.2913505}\BibitemShut
  {NoStop}%
\bibitem [{\citenamefont {{Zemanov{\'{a}} Die{\v{s}}kov{\'{a}}}}\ \emph
  {et~al.}(2013)\citenamefont {{Zemanov{\'{a}} Die{\v{s}}kov{\'{a}}}},
  \citenamefont {Ferretti},\ and\ \citenamefont
  {Bokes}}]{ZemanovaDieskova2013}%
  \BibitemOpen
  \bibfield  {author} {\bibinfo {author} {\bibfnamefont {M.}~\bibnamefont
  {{Zemanov{\'{a}} Die{\v{s}}kov{\'{a}}}}}, \bibinfo {author} {\bibfnamefont
  {A.}~\bibnamefont {Ferretti}}, \ and\ \bibinfo {author} {\bibfnamefont
  {P.}~\bibnamefont {Bokes}},\ }\href {\doibase 10.1103/PhysRevB.87.195107}
  {\bibfield  {journal} {\bibinfo  {journal} {Physical Review B}\ }\textbf
  {\bibinfo {volume} {87}},\ \bibinfo {pages} {195107} (\bibinfo {year}
  {2013})}\BibitemShut {NoStop}%
\bibitem [{\citenamefont {Jung}\ \emph {et~al.}(2009)\citenamefont {Jung},
  \citenamefont {Kim}, \citenamefont {Jung}, \citenamefont {Im}, \citenamefont
  {Pashkin}, \citenamefont {Astafiev}, \citenamefont {Nakamura}, \citenamefont
  {Lee}, \citenamefont {Miyamoto},\ and\ \citenamefont {Tsai}}]{Jung2009}%
  \BibitemOpen
  \bibfield  {author} {\bibinfo {author} {\bibfnamefont {H.}~\bibnamefont
  {Jung}}, \bibinfo {author} {\bibfnamefont {Y.}~\bibnamefont {Kim}}, \bibinfo
  {author} {\bibfnamefont {K.}~\bibnamefont {Jung}}, \bibinfo {author}
  {\bibfnamefont {H.}~\bibnamefont {Im}}, \bibinfo {author} {\bibfnamefont
  {Y.~A.}\ \bibnamefont {Pashkin}}, \bibinfo {author} {\bibfnamefont
  {O.}~\bibnamefont {Astafiev}}, \bibinfo {author} {\bibfnamefont
  {Y.}~\bibnamefont {Nakamura}}, \bibinfo {author} {\bibfnamefont
  {H.}~\bibnamefont {Lee}}, \bibinfo {author} {\bibfnamefont {Y.}~\bibnamefont
  {Miyamoto}}, \ and\ \bibinfo {author} {\bibfnamefont {J.~S.}\ \bibnamefont
  {Tsai}},\ }\href {\doibase 10.1103/PhysRevB.80.125413} {\bibfield  {journal}
  {\bibinfo  {journal} {Physical Review B}\ }\textbf {\bibinfo {volume} {80}},\
  \bibinfo {pages} {125413} (\bibinfo {year} {2009})}\BibitemShut {NoStop}%
\bibitem [{\citenamefont {Cyster}\ \emph {et~al.}(2020)\citenamefont {Cyster},
  \citenamefont {Smith}, \citenamefont {Vaitkus}, \citenamefont {Vogt},
  \citenamefont {Russo},\ and\ \citenamefont {Cole}}]{Cyster2020}%
  \BibitemOpen
  \bibfield  {author} {\bibinfo {author} {\bibfnamefont {M.~J.}\ \bibnamefont
  {Cyster}}, \bibinfo {author} {\bibfnamefont {J.~S.}\ \bibnamefont {Smith}},
  \bibinfo {author} {\bibfnamefont {J.~A.}\ \bibnamefont {Vaitkus}}, \bibinfo
  {author} {\bibfnamefont {N.}~\bibnamefont {Vogt}}, \bibinfo {author}
  {\bibfnamefont {S.~P.}\ \bibnamefont {Russo}}, \ and\ \bibinfo {author}
  {\bibfnamefont {J.~H.}\ \bibnamefont {Cole}},\ }\href {\doibase
  10.1103/physrevresearch.2.013110} {\bibfield  {journal} {\bibinfo  {journal}
  {Physical Review Research}\ }\textbf {\bibinfo {volume} {2}},\ \bibinfo
  {pages} {013110} (\bibinfo {year} {2020})},\ \Eprint
  {http://arxiv.org/abs/1905.12214} {arXiv:1905.12214} \BibitemShut {NoStop}%
\bibitem [{\citenamefont {Curran}\ \emph {et~al.}(1982)\citenamefont {Curran},
  \citenamefont {Page},\ and\ \citenamefont {Pick}}]{Curran1982}%
  \BibitemOpen
  \bibfield  {author} {\bibinfo {author} {\bibfnamefont {J.~E.}\ \bibnamefont
  {Curran}}, \bibinfo {author} {\bibfnamefont {J.~S.}\ \bibnamefont {Page}}, \
  and\ \bibinfo {author} {\bibfnamefont {U.}~\bibnamefont {Pick}},\ }\href
  {\doibase 10.1016/0040-6090(82)90460-6} {\bibfield  {journal} {\bibinfo
  {journal} {Thin Solid Films}\ }\textbf {\bibinfo {volume} {97}},\ \bibinfo
  {pages} {259} (\bibinfo {year} {1982})}\BibitemShut {NoStop}%
\bibitem [{\citenamefont {Sullivan}\ \emph {et~al.}(1998)\citenamefont
  {Sullivan}, \citenamefont {Barbour}, \citenamefont {Dunn}, \citenamefont
  {Son}, \citenamefont {Montes}, \citenamefont {Missert},\ and\ \citenamefont
  {Copeland}}]{Sullivan1998a}%
  \BibitemOpen
  \bibfield  {author} {\bibinfo {author} {\bibfnamefont {J.~P.}\ \bibnamefont
  {Sullivan}}, \bibinfo {author} {\bibfnamefont {J.~C.}\ \bibnamefont
  {Barbour}}, \bibinfo {author} {\bibfnamefont {R.~G.}\ \bibnamefont {Dunn}},
  \bibinfo {author} {\bibfnamefont {K.~A.}\ \bibnamefont {Son}}, \bibinfo
  {author} {\bibfnamefont {L.~P.}\ \bibnamefont {Montes}}, \bibinfo {author}
  {\bibfnamefont {N.}~\bibnamefont {Missert}}, \ and\ \bibinfo {author}
  {\bibfnamefont {R.~G.}\ \bibnamefont {Copeland}},\ }\href
  {http://www.osti.gov/scitech/servlets/purl/1916} {\bibfield  {journal}
  {\bibinfo  {journal} {The Electrochemical Society Meeting}\ } (\bibinfo
  {year} {1998})}\BibitemShut {NoStop}%
\bibitem [{\citenamefont {Koski}\ \emph {et~al.}(1999)\citenamefont {Koski},
  \citenamefont {H{\"{o}}ls{\"{a}}},\ and\ \citenamefont {Juliet}}]{Koski1999}%
  \BibitemOpen
  \bibfield  {author} {\bibinfo {author} {\bibfnamefont {K.}~\bibnamefont
  {Koski}}, \bibinfo {author} {\bibfnamefont {J.}~\bibnamefont
  {H{\"{o}}ls{\"{a}}}}, \ and\ \bibinfo {author} {\bibfnamefont
  {P.}~\bibnamefont {Juliet}},\ }\href {\doibase 10.1016/S0040-6090(98)01232-2}
  {\bibfield  {journal} {\bibinfo  {journal} {Thin Solid Films}\ }\textbf
  {\bibinfo {volume} {339}},\ \bibinfo {pages} {240} (\bibinfo {year}
  {1999})}\BibitemShut {NoStop}%
\bibitem [{\citenamefont {Trybula}\ and\ \citenamefont
  {Korzhavyi}(2019)}]{Trybula2019}%
  \BibitemOpen
  \bibfield  {author} {\bibinfo {author} {\bibfnamefont {M.~E.}\ \bibnamefont
  {Trybula}}\ and\ \bibinfo {author} {\bibfnamefont {P.~A.}\ \bibnamefont
  {Korzhavyi}},\ }\href {\doibase 10.1021/acs.jpcc.8b06910} {\bibfield
  {journal} {\bibinfo  {journal} {Journal of Physical Chemistry C}\ }\textbf
  {\bibinfo {volume} {123}},\ \bibinfo {pages} {334} (\bibinfo {year}
  {2019})}\BibitemShut {NoStop}%
\bibitem [{\citenamefont {Evangelisti}\ \emph {et~al.}(2017)\citenamefont
  {Evangelisti}, \citenamefont {Stiefel}, \citenamefont {Guseva}, \citenamefont
  {{Partovi Nia}}, \citenamefont {Hauert}, \citenamefont {Hack}, \citenamefont
  {Jeurgens}, \citenamefont {Ambrosio}, \citenamefont {Pasquarello},
  \citenamefont {Schmutz},\ and\ \citenamefont
  {Cancellieri}}]{Evangelisti2017}%
  \BibitemOpen
  \bibfield  {author} {\bibinfo {author} {\bibfnamefont {F.}~\bibnamefont
  {Evangelisti}}, \bibinfo {author} {\bibfnamefont {M.}~\bibnamefont
  {Stiefel}}, \bibinfo {author} {\bibfnamefont {O.}~\bibnamefont {Guseva}},
  \bibinfo {author} {\bibfnamefont {R.}~\bibnamefont {{Partovi Nia}}}, \bibinfo
  {author} {\bibfnamefont {R.}~\bibnamefont {Hauert}}, \bibinfo {author}
  {\bibfnamefont {E.}~\bibnamefont {Hack}}, \bibinfo {author} {\bibfnamefont
  {L.~P.}\ \bibnamefont {Jeurgens}}, \bibinfo {author} {\bibfnamefont
  {F.}~\bibnamefont {Ambrosio}}, \bibinfo {author} {\bibfnamefont
  {A.}~\bibnamefont {Pasquarello}}, \bibinfo {author} {\bibfnamefont
  {P.}~\bibnamefont {Schmutz}}, \ and\ \bibinfo {author} {\bibfnamefont
  {C.}~\bibnamefont {Cancellieri}},\ }\href {\doibase
  10.1016/j.electacta.2016.12.090} {\bibfield  {journal} {\bibinfo  {journal}
  {Electrochimica Acta}\ }\textbf {\bibinfo {volume} {224}},\ \bibinfo {pages}
  {503} (\bibinfo {year} {2017})}\BibitemShut {NoStop}%
\bibitem [{\citenamefont {Zeng}\ \emph
  {et~al.}(2015{\natexlab{b}})\citenamefont {Zeng}, \citenamefont {Krantz},
  \citenamefont {Nik}, \citenamefont {Delsing},\ and\ \citenamefont
  {Olsson}}]{Zeng2015}%
  \BibitemOpen
  \bibfield  {author} {\bibinfo {author} {\bibfnamefont {L.~J.}\ \bibnamefont
  {Zeng}}, \bibinfo {author} {\bibfnamefont {P.}~\bibnamefont {Krantz}},
  \bibinfo {author} {\bibfnamefont {S.}~\bibnamefont {Nik}}, \bibinfo {author}
  {\bibfnamefont {P.}~\bibnamefont {Delsing}}, \ and\ \bibinfo {author}
  {\bibfnamefont {E.}~\bibnamefont {Olsson}},\ }\href {\doibase
  10.1063/1.4919224} {\bibfield  {journal} {\bibinfo  {journal} {Journal of
  Applied Physics}\ }\textbf {\bibinfo {volume} {117}},\ \bibinfo {pages}
  {163915} (\bibinfo {year} {2015}{\natexlab{b}})}\BibitemShut {NoStop}%
\bibitem [{\citenamefont {Kohlstedt}\ \emph {et~al.}(1993)\citenamefont
  {Kohlstedt}, \citenamefont {Hallmanns}, \citenamefont {Nevirkovets},
  \citenamefont {Guggi},\ and\ \citenamefont {Heiden}}]{Kohlstedt1993}%
  \BibitemOpen
  \bibfield  {author} {\bibinfo {author} {\bibfnamefont {H.}~\bibnamefont
  {Kohlstedt}}, \bibinfo {author} {\bibfnamefont {G.}~\bibnamefont
  {Hallmanns}}, \bibinfo {author} {\bibfnamefont {I.}~\bibnamefont
  {Nevirkovets}}, \bibinfo {author} {\bibfnamefont {D.}~\bibnamefont {Guggi}},
  \ and\ \bibinfo {author} {\bibfnamefont {C.}~\bibnamefont {Heiden}},\ }\href
  {http://ieeexplore.ieee.org/xpls/abs_all.jsp?arnumber=233939} {\bibfield
  {journal} {\bibinfo  {journal} {IEEE Transactions on Applied
  Superconductivity}\ }\textbf {\bibinfo {volume} {3}},\ \bibinfo {pages}
  {2197} (\bibinfo {year} {1993})}\BibitemShut {NoStop}%
\bibitem [{\citenamefont {Zhu}\ and\ \citenamefont {Park}(2006)}]{Zhu2006}%
  \BibitemOpen
  \bibfield  {author} {\bibinfo {author} {\bibfnamefont {J.-G.}\ \bibnamefont
  {Zhu}}\ and\ \bibinfo {author} {\bibfnamefont {C.}~\bibnamefont {Park}},\
  }\href {\doibase 10.1016/S1369-7021(06)71693-5} {\bibfield  {journal}
  {\bibinfo  {journal} {Materials Today}\ }\textbf {\bibinfo {volume} {9}},\
  \bibinfo {pages} {36} (\bibinfo {year} {2006})}\BibitemShut {NoStop}%
\bibitem [{\citenamefont {Gradshteĭn}\ \emph {et~al.}(2007)\citenamefont
  {Gradshteĭn}, \citenamefont {Ryzhik},\ and\ \citenamefont
  {Jeffrey}}]{Gradshtein2007}%
  \BibitemOpen
  \bibfield  {author} {\bibinfo {author} {\bibfnamefont {I.~S.}\ \bibnamefont
  {Gradshteĭn}}, \bibinfo {author} {\bibfnamefont {I.~M.}\ \bibnamefont
  {Ryzhik}}, \ and\ \bibinfo {author} {\bibfnamefont {A.}~\bibnamefont
  {Jeffrey}},\ }\href@noop {} {\emph {\bibinfo {title} {{Table of integrals,
  series, and products}}}},\ \bibinfo {edition} {7th}\ ed.\ (\bibinfo
  {publisher} {Academic Press},\ \bibinfo {address} {Amsterdam ; Boston},\
  \bibinfo {year} {2007})\BibitemShut {NoStop}%
\bibitem [{\citenamefont {Gale}(1997)}]{Gale1997}%
  \BibitemOpen
  \bibfield  {author} {\bibinfo {author} {\bibfnamefont {J.}~\bibnamefont
  {Gale}},\ }\href {\doibase 10.1039/A606455H} {\bibfield  {journal} {\bibinfo
  {journal} {Journal of the Chemical Society: Faraday Transactions}\ }\textbf
  {\bibinfo {volume} {93}},\ \bibinfo {pages} {629} (\bibinfo {year}
  {1997})}\BibitemShut {NoStop}%
\bibitem [{\citenamefont {Holian}\ \emph {et~al.}(1995)\citenamefont {Holian},
  \citenamefont {Voter},\ and\ \citenamefont {Ravelo}}]{Holian1995}%
  \BibitemOpen
  \bibfield  {author} {\bibinfo {author} {\bibfnamefont {B.~L.}\ \bibnamefont
  {Holian}}, \bibinfo {author} {\bibfnamefont {A.~F.}\ \bibnamefont {Voter}}, \
  and\ \bibinfo {author} {\bibfnamefont {R.}~\bibnamefont {Ravelo}},\ }\href
  {\doibase 10.1103/PhysRevE.52.2338} {\bibfield  {journal} {\bibinfo
  {journal} {Physical Review E}\ }\textbf {\bibinfo {volume} {52}},\ \bibinfo
  {pages} {2338} (\bibinfo {year} {1995})}\BibitemShut {NoStop}%
\bibitem [{Note1()}]{Note1}%
  \BibitemOpen
  \bibinfo {note} {A non-linear least-square method with the bi-square robust
  fitting option was used with the curve fitting toolbox.}\BibitemShut {Stop}%
\end{thebibliography}%

\appendix

\section{Estimating the frequency of gas--surface interactions}
\label{appendix:gas_surface}
% !TEX root = deposition.tex

Here we aim to calculate the number of atoms in the diffuse oxygen gas which will strike an area the size of our simulated surface.
The hypothesis is that the frequency of atom strikes on the surface is low enough that it is a good approximation to consider them as independent events.
We are interested in the flux of atoms striking a small surface region $dA$ over time
\begin{equation}
    d\Phi = \frac{dN}{dA\,dt}
    \label{eqn:atomic_flux}
\end{equation}

We start by considering an arbitrary velocity distribution for the particles in the gas. This can be written as 
\begin{equation}
    G(v_x, v_y, v_z).
\end{equation}
The number of atoms inside an infinitesimal part of the velocity space can then be written as
\begin{equation}
    dN = N \, G(v_x, v_y, v_z) \, dv_x \, dv_y \, dv_z
\end{equation}
where $N$ is the total number of particles.
By assuming a spherical symmetry and transforming to spherical coordinates we can equivalently write
\begin{equation}
    dN = N \, G(v) \, v^2 dv \, \sin \theta \, d\theta \, d\phi
    \label{eqn:dn_spherical}
\end{equation}
where $G(v)$ is defined only by the particle's speed $v$.

\begin{figure}
    \includegraphics[width=2.05 in]{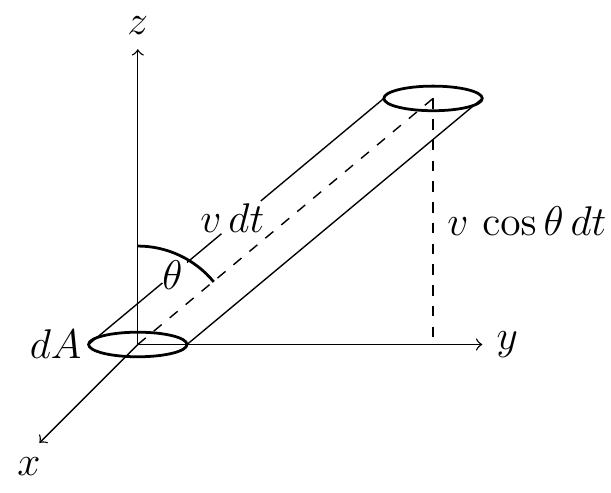}
    \caption
    {
        A diagrammatic representation of the volume $dV$ as defined in Eq.~\ref{eqn:infinitesimal_volume}.
        Conceptually this is a slant cylinder containing all of the particles at a velocity $v$ which will reach the surface $dA$ after a given time $dt$.
        The angle of the cylinder relative to the surface is defined as $\theta$.
    }
    \label{fig:flux_particle_cylinder}
\end{figure}

In real space, we consider only those atoms approaching the surface from a particular direction defined by $\theta$ and $\phi$. 
We narrow this definition to include only particles at a given velocity $v$. 
These particles are at a distance $v\,dt$ from the surface. 
Together these quantities define an infinitesimal volume $dV$ (depicted in Fig.~\ref{fig:flux_particle_cylinder}):
\begin{equation}
    dV = v \cos \theta \ dt \, dA
    \label{eqn:infinitesimal_volume}
\end{equation}
where $dA$ is the surface element. 
The number of atoms inside the volume $dV$ can then be calculated from the concentration of atoms $n$:
\begin{equation}
    N_V = n \, dV.
    \label{eqn:num_particles_in_cylinder}
\end{equation}
By substituting this in Eq.~\ref{eqn:dn_spherical} with $N \rightarrow N_V$ we obtain
\begin{equation}
    dN = n \, v \cos\theta \, dt \, dA \, G(v) \, v^2 \, dv \, \sin \theta \, d\theta \, d\phi
\end{equation}
Returning to the expression for the atomic flux (Eq.~\ref{eqn:atomic_flux}) we have
\begin{align}
    d\Phi = \frac{dN}{dA \, dt} &= \frac{n \, v \cos\theta \, dt \, dA \, G(v) \, v^2 \, dv \, \sin \theta \, d\theta \, d\phi}{dA \, dt} \\
    &= n \, v \cos\theta \, G(v) \, v^2 \, dv \, \sin \theta \, d\theta \, d\phi.
\end{align}

Finally we are able to integrate this equation over a hemisphere to account for all the incoming particles on one side of the plane:
\begin{align}
    \Phi &= n
        \left[ \int_0^{\infty} v^3 \, G(v) \, dv  \right] 
        \left[ \int_0^{\pi/2} \sin\theta \cos\theta\ d\theta  \right]
        \left[ \int_0^{2\pi} d\phi \right] \\
    &= n
        \left[ \int_0^{\infty} v^3 \, G(v) \, dv  \right] 
        \left[ 1 / 2 \right]
        \left[ 2\pi \right] \\
    &= n \pi \left[ \int_0^{\infty} v^3 \, G(v) \, dv  \right]  \\
    &= \frac{n}{4} \left[ \int_0^{\infty} v \, f(v) \, dv  \right]
\end{align}
where we have introduced the function $f(v)$
\begin{equation}
f(v) \equiv 4\pi v^2 G(v).
\end{equation}
We define the average velocity of a particle in the distribution $f(v)$ as
\begin{equation}
    \bar{v} \equiv \int_0^{\infty} v\,f(v)\ dv 
\end{equation}
which allows us to express the flux of the particles simply as
\begin{equation}
    \Phi = \frac{n\bar{v}}{4}.
    \label{eqn:flux_ave_vel}
\end{equation}

Due to the low pressure state of the system we consider the gas to obey the ideal gas law equation.
Under these conditions the Maxwell--Boltzmann velocity distribution is well suited to describing the statistics of the particles and the function $f(v)$ has the form
\begin{equation} 
    f(v) = \left( \frac{m}{2\pi k_B T} \right)^{3/2} \ 4\pi v^2\ \exp \left( - \frac{mv^2}{2 k_B T} \right)
\end{equation}
Therefore
\begin{align}
    \bar{v} &= \int_0^{\infty} v\,f(v)\ dv \\
    &= \left( \frac{m}{2\pi k_B T} \right)^{3/2} \ 4\pi \int_0^{\infty} v^3\ \exp \left( - \frac{mv^2}{2 k_B T} \right)\ dv
\end{align}
Consulting a table of integrals (Ref.~\citenum{Gradshtein2007}, p.360; 3.461, Eq.~2) we find:
\begin{equation}
    \int_0^{\infty} x^{2n+1} e^{-px^2} \ dx  = \frac{n!}{2p^{n+1}} \quad\forall\quad p > 0.
\end{equation}
Substituting $n \rightarrow 1$, $p \rightarrow {m}/{2k_B T}$ and $x \rightarrow v$ gives
\begin{equation}
    \bar{v} = \left( \frac{m}{2\pi k_B T} \right)^{3/2} \ 2\pi \ \left( \frac{2 k_B T}{m} \right)^2 = \sqrt{\frac{8 k_B T}{\pi m}}.
\end{equation}
This is the average velocity of a particle of a given mass $m$ in a Maxwell--Boltzmann distribution with the temperature $T$.

We can also rewrite the concentration $n$ in terms of the gas pressure using the ideal gas law
 \begin{equation}
    n = \frac{N}{V} = \frac{p}{k_B T}.
\end{equation}
Substituting these expressions for $\bar{v}$ and $n$ in Eq.~\ref{eqn:flux_ave_vel} we obtain the flux as a function of the pressure, temperature and the mass of the particles in the gas:
\begin{equation}
    \Phi(p, m, T) = \frac{p}{ 4 k_B T} \, \sqrt{\frac{8 k_B T}{\pi m}} = \frac{p}{\sqrt{ 2\pi m k_B T }}.
\end{equation}

Multiplying the flux by the area of the surface in the simulation allows us to estimate the number of atoms which strike the surface per unit time.
Considering as an example the oxygen pressure of 1.33 $\times$ 10$^{-4}$ Pa reported by Jeurgens et~al.\ and a temperature of 300 K gives a rate of approximately 52 surface interactions per second on a surface with area $A \simeq$ 32.0 $\times$ 32.0 \AA$^2$ \cite{Jeurgens2002c}.
This is equivalent to one of the largest structures we simulate.
The limitation on the size of the surface arises from our choice of simulation package, the complexity of the empirical potential, and the available computing power on current supercomputing facilities.
From our estimate one atom interacts with the surface about every 20~ms.
As the total simulation time for one oxygen atom to be deposited on the surface is of the order of tens of picoseconds, it is a good approximation to consider these atom strikes as independent events.

\section{Magnitude of the coupling strength for the Nos\`e--Hoover thermostat}
\label{appendix:thermostat}
% !TEX root = deposition.tex

As the molecular dynamics simulations are performed at a constant temperature a thermostatting process must be included to describe heat transfer in and out of the simulation cell.
We use the Nos\`e--Hoover thermostat algorithm in the simulation package GULP to perform simulations in the canonical (or NVT) ensemble.\cite{Gale1997,Hoover1985}
This algorithm maintains a constant temperature in the simulation by coupling the particles to a fictional external thermal reservoir.

To understand how this works mathematically, consider the relationship between the instantaneous kinetic temperature $T$ of the simulation and the velocities of the simulated particles.
By the equipartition theorem we equate the thermal energy with the sum of the kinetic energies of the particles:
\begin{equation}
    d N k_B T = \sum_{i=1}^{N} m_i {\mathbf{v}_i}^2
\end{equation}
In this expression $d$ is the number of dimensions, $N$ is the number of particles and $k_B$ is Boltzmann's constant. The atomic mass and velocity of particle $i$ are given by $m_i$ and $\mathbf{v}_i$ respectively. 

The thermostat introduces two new quantities: a coupling constant $\nu$ and a heat flow variable $\xi$ which is related to the effective ``mass'' of the fictional heat reservoir.\cite{Nose1984}
The equations of motion are modified for particle $i$ such that
\begin{align}
    m_i \mathbf{a}_i &= \mathbf{F}_{i} - \nu \xi m_i \mathbf{v}_i 
    \label{eqn:modified_acceleration} \\
    \frac{d\xi}{dt} &= \nu \left( \frac{T}{T_0} - 1 \right)
    \label{eqn:dxi_dt}
\end{align}
where $\mathbf{F}_{i}$ is the force on particle $i$ and $T_0$ is the target temperature for the thermostat, i.e. the desired temperature for the simulation.
The relationship between $\xi$ and $\nu$ arises from a consideration of the canonical NVT dynamics the thermostat is designed to reproduce.\cite{Hoover1985}
When the instantaneous temperature $T$ exceeds the target temperature, the value of Eq.~\ref{eqn:dxi_dt} is positive. 
This means that the value of $\xi$ will increase in Eq.~\ref{eqn:modified_acceleration}, arresting the acceleration of the particle.
By modifying the acceleration of the particles, heat is effectively added and removed from the system.
The total energy of the system is given by the sum of the kinetic energy $K$, the potential energy $U$, and a contribution from the heat flow variable $\xi$:
\begin{equation}
    E(\mathbf{q}, \mathbf{p}, \xi) = K(\mathbf{q}) + U(\mathbf{p}) + \frac{d}{2} N k_B T_0 \xi^2
\end{equation}
where $\mathbf{q}$ are the positions and $\mathbf{p}$ are the momenta of the particles.
The thermostat coupling parameter $\nu$ defines the strength of the coupling, i.e. how quickly the particle velocities will respond to a temperature either above or below the target temperature. 
Here we establish a method for determining the appropriate strength of this coupling so as to accurately describe a canonical ensemble.

\begin{figure}
    \centering
    \includegraphics{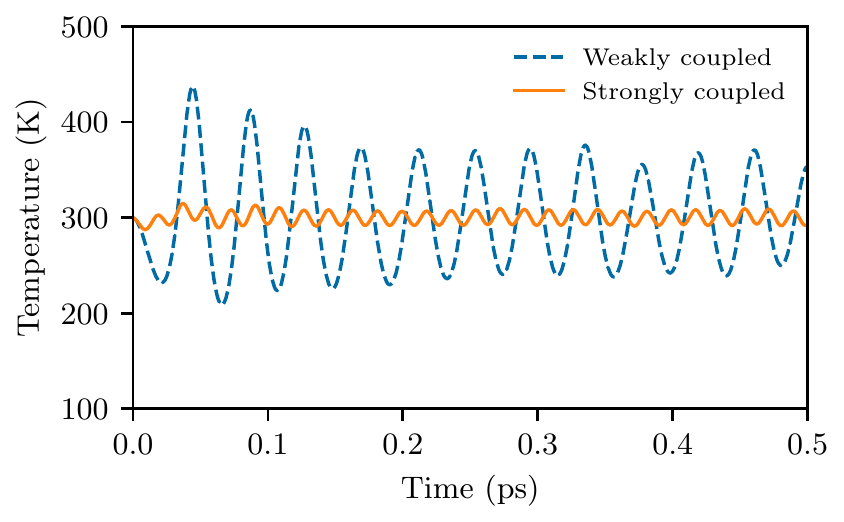}
    \caption [The effect of the thermostat coupling on temperature oscillations] 
    {
        Oscillations in the temperature of the simulation are dependent on the strength of the Nos\`e--Hoover thermostat coupling parameter $\nu$. 
        The data displayed is for a periodic aluminium supercell containing 256 atoms. 
    }
    \label{fig:temperature_oscillations}
\end{figure}

While the Nos\`e--Hoover thermostat equations were originally designed to reproduce the statistics of the canonical ensemble, Holian et al.\ have shown that using an incorrect value for the coupling can produce non-physical oscillations in the temperature.\cite{Holian1995}
They present a number of physically motivated approaches for setting the thermostat coupling, one of which is to consider statistical fluctuations in the temperature over the course of the simulation.
When the particles are strongly coupled to the external heat bath temperature fluctuations about the mean are small.
Conversely, weak coupling allows the simulated system to act independently of the heat bath, leading to large fluctuations.
The dependence of the temperature oscillations on $\nu$ is demonstrated in Fig.~\ref{fig:temperature_oscillations}. 
In the weakly coupled case the temperature varies quite dramatically about the mean; as much as $\pm$100 K. 
The strongly coupled case constrains the temperature more closely to the mean value.

For a canonical (or NVT) system Holian et~al.\ give an equation which predicts the variance of the temperature:
\begin{equation}
    \sigma^{2}_{NVT}(d,\bar{T},N) = \frac{2}{d} \frac{ \bar{T}}{N}^2
    \label{eqn:holian_canonical_variance}
\end{equation}
where $d$ is the number of dimensions, $\bar{T}$ is the mean temperature and $N$ is the number of particles. 
By calculating the variance in the temperature we obtain a metric which represents the magnitude of the fluctuations.

\begin{figure}[b]
    \centering
    \includegraphics{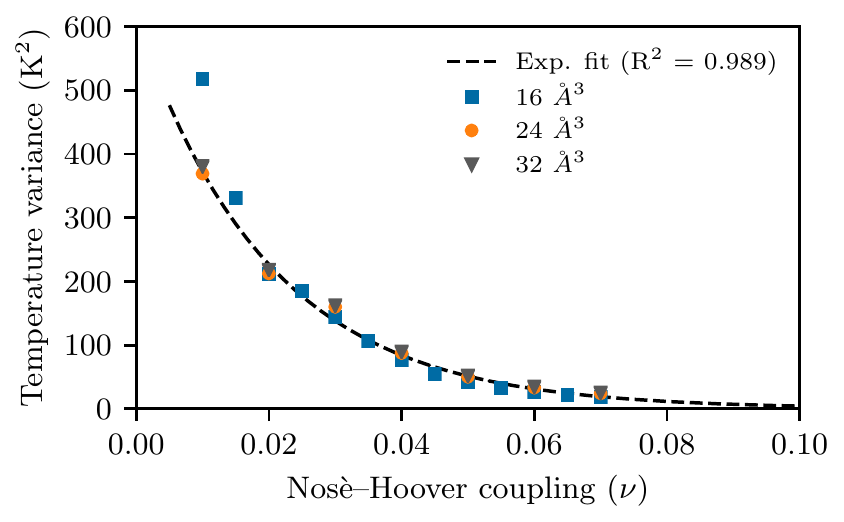}
    \caption [Temperature variance as a function of the Nos\`e--Hoover coupling] 
    {
        The calculated variance in the temperature as the Nos\`e--Hoover thermostat coupling $\nu$ is varied. 
        As expected the temperature oscillates more about the mean for small values of $\nu$, where the thermostat coupling is weak. 
        The dashed grey line shows the exponential function fitted to the data.
    }
    \label{fig:nose_hoover_variance}
\end{figure}

In order to study how the temperature variance changes as a function of $\nu$, we simulate aluminium supercells with a range of coupling strengths.
Three cubic supercells are simulated with side lengths of 16~\AA, 24~\AA, and 32~\AA.
Each supercell is simulated with periodic boundary conditions for a total of approximately 0.8~ns solving the equations of motion every 1~fs.
The total duration of $\sim$0.8~ns is reached by running a series of independent 10~ps calculations where the initial velocities of the particles in each calculation are randomised.
Running a large number of independent simulations allows a reliable estimate to be obtained from the statistical analysis.
The temperature variance was calculated for each 10.0~ps simulation before being averaged to generate the data on Fig.~\ref{fig:nose_hoover_variance}.
An exponential fit to the data (R$^2$ = 0.9897) was calculated with MATLAB\footnote{
    A non-linear least-square method with the bi-square robust fitting option was used with the curve fitting toolbox.
}
\begin{equation}
    \sigma^{2}_{T}(\nu) = a \exp{(b \nu)}
    \label{eqn:exponential_fit}
\end{equation}
where the constants were found to be $a = $610 and $b = -$49.6.
Using the relationship between the coupling parameter and the variance of the temperature we are able to set the coupling to obtain the canonical temperature variance.

To demonstrate how $\nu$ is chosen for a given simulation with this expression, we consider as an example a typical simulation of 1000 particles where the mean temperature is $\bar{T} = 300$~K.
Using Eq.~\ref{eqn:holian_canonical_variance} we calculate the expected canonical variance $\sigma^{2}_{NVT}$ as a function of the temperature and number of particles. 
\begin{equation}
    \sigma^{2}_{NVT}(d, \bar{T}, N) = \frac{2}{d} \frac{\bar{T}}{N}^2 = \frac{2}{3} \frac{(300~\mathrm{K})}{1000}^2 = 60~\mathrm{K}^2
\end{equation}
To find the appropriate value of $\nu$ for this simulation we invert Eq.~\ref{eqn:exponential_fit} and express it in terms of the variance:
\begin{equation}
    \nu = \frac{\ln{(\sigma^{2}_{NVT})} - \ln{(a)}}{b}
    \label{eqn:inverted_fit}
\end{equation}
Substituting the appropriate values into Eq.~\ref{eqn:inverted_fit} then gives
\begin{equation}
    \nu = \frac{\ln{(60~\mathrm{K}^2)} - \ln{(610)}}{-49.6} = 0.0468
\end{equation}
In other words, in a simulation of 1000 particles at 300~K the canonical variance is reproduced by setting the thermostat coupling to $\nu =$ 0.0468.

From a statistical analysis of the simulation temperature over a large period of time for different values of $\nu$ an exponential relationship is found between the temperature variance and the thermostat coupling.
This allows us to express the coupling strength $\nu$ as a function of the variance (Eq.~\ref{eqn:exponential_fit}).
Using the expression for the temperature variance in a canonical ensemble (Eq.~\ref{eqn:holian_canonical_variance}) we are able to calculate the expected variance for any given system.\cite{Holian1995}
For our molecular dynamics simulations, we determine the expected variance as per Holian et~al.\ and use Eq.~\ref{eqn:inverted_fit} to set the value of $\nu$ appropriately.

\end{document}